\documentclass{article}
\pdfoutput=1
\usepackage{jinstpub}
\usepackage{multirow}
\usepackage{array}
\newcommand{\PreserveBackslash}[1]{\let\temp=\\#1\let\\=\temp}
\newcolumntype{C}[1]{>{\PreserveBackslash\centering}p{#1}}
\usepackage[utf8]{inputenc}

\author[a]{M. Dyndal}
\author[a]{V. Dao} 
\author[i]{P. Allport}
\author[a,b]{I. Asensi Tortajada}
\author[m]{M. Barbero}
\author[m]{S. Bhat}
\author[c]{D. Bortoletto}
\author[a,j]{I. Berdalovic}
\author[k]{C. Bespin}
\author[f]{C. Buttar}
\author[k]{I. Caicedo}
\author[a,d]{R. Cardella}
\author[a,e]{F. Dachs}
\author[n]{Y. Degerli}
\author[l]{H. Denizli}
\author[a,f]{L. Flores Sanz de Acedo} 
\author[i]{P. Freeman}  
\author[i]{L. Gonella}
\author[m]{A. Habib}
\author[k]{T. Hemperek}
\author[k]{T. Hirono}
\author[g]{B. Hiti} 
\author[a]{T. Kugathasan}
\author[g]{I. Mandi\'c}
\author[f]{D. Maneuski}
\author[g]{M. Miku\v{z}}
\author[k]{K. Moustakas}
\author[a]{M. Munker} 
\author[l]{K. Y. Oyulmaz} 
\author[m]{P. Pangaud}
\author[a]{H. Pernegger} 
\author[a]{F. Piro} 
\author[a]{P. Riedler} 
\author[d]{H. Sandaker}
\author[a]{E. J. Schioppa} 
\author[n]{P. Schwemling}
\author[a,c]{A. Sharma}
\author[f]{L. Simon Argemi}
\author[a]{C. Solans Sanchez}
\author[a]{W. Snoeys} 
\author[j]{T. Suligoj} 
\author[k]{T. Wang}
\author[k]{N. Wermes} 
\author[i,o]{S. Worm}

\affiliation[a]{CERN, Geneva, Switzerland}
\affiliation[b]{University of Valencia and Consejo Superior de Investigaci Científicas (CSIC), Valencia, Spain}
\affiliation[c]{University of Oxford, Oxford, United Kingdom}
\affiliation[d]{University of Oslo, Oslo, Norway}
\affiliation[e]{Vienna University of Technology, Vienna, Austria}
\affiliation[f]{University of Glasgow, Glasgow, United Kingdom}
\affiliation[g]{Jo\v{z}ef Stefan Institute, Ljubljana, Slovenia}
\affiliation[i]{University of Birmingham, Birmingham, United Kingdom}
\affiliation[j]{University of Zagreb, Zagreb, Croatia}
\affiliation[k]{Rheinische Friedrich-Wilhelms Universität Bonn, Bonn, Germany}
\affiliation[l]{Bolu Abant Izzet Baysal University, Bolu, Turkey}
\affiliation[m]{Aix Marseille University, CNRS/IN2P3, CPPM, Marseille, France}
\affiliation[n]{CEA-IRFU, Paris, France}
\affiliation[o]{Deutsches Elektron-Synchrotron DESY, Hamburg, Germany}

\emailAdd{heinz.pernegger@cern.ch}

\title{Mini-MALTA: Radiation hard pixel designs for small-electrode monolithic CMOS sensors for the High Luminosity LHC}

\abstract{Depleted Monolithic Active Pixel Sensor (DMAPS) prototypes developed in the TowerJazz 180 nm CMOS imaging process have been designed in the context of the ATLAS upgrade Phase-II at the HL-LHC. The pixel sensors are characterized by a small collection electrode (3 $\mu$m) to minimize capacitance, a small pixel size ($36.4\times 36.4$ $\mu$m$^2$), and are produced on high resistivity epitaxial p-type silicon. The design targets a radiation hardness of $1\times10^{15}$ 1 MeV n$_{eq}$/cm$^{2}$, compatible with the outermost layer of the ATLAS ITK Pixel detector. This paper presents the results from characterization in particle beam tests of the Mini-MALTA prototype that implements a mask change or an additional implant to address the inefficiencies on the pixel edges. Results show full efficiency after a dose of  $1\times10^{15}$ 1 MeV n$_{eq}$/cm$^{2}$.}

\date{September 2019}

\notoc

\begin{document}

\maketitle

\section{Introduction}
Depleted Monolithic Active Pixel Sensor (DMAPS) prototypes have been developed in the TowerJazz 180~nm CMOS imaging process with the aim to explore their viability for the Phase-II upgrade of ATLAS for the High Luminosity LHC ~\cite{ItkPixelTdr}, and for future HEP experiments \cite{Investigator}\cite{Snoeys:2017hjn}\cite{Wang2018}. With previous developments focusing on low-radiation environments \cite{Algieri:2013}, special interest lies now on the radiation hardness of this technology up to 100 Mrad in Total Ionizing Dose (TID) and $\ge 1\times10^{15}$ 1 MeV n$_{eq}$/cm$^{2}$ in Non-Ionizing Energy Loss (NIEL) in order to be used in the harsh environment of these experiments. Monolithic CMOS sensors additionally allow to minimize scattering material for best tracking performance. The developments reported here investigate pixel sensors, dubbed MALTA sensor and Mini-MALTA sensor, with small electrodes (electrode diameter 3~$\mu$m at 36.4 $\mu$m pixel pitch). 

The advantage of small collection electrodes lies in the resulting small capacitance, which in turn helps to minimize noise and achieve low power dissipation in the active area. However, detection efficiency after irradiation in sensors with small electrodes can be critically affected in the pixel corners \cite{Cardella:2019ksc}\cite{Caicedo2019}. To improve this, we designed special p-type and n-type implant geometries \cite{Munker:2019vdo}. These special implant geometries are implemented in sub-matrices of the Mini-MALTA sensor. The detection efficiency for different pixel designs before and after neutron irradiation are studied in beam tests to determine the optimal pixel design for radiation hard depleted monolithic CMOS sensors.


\section{Sensors with small collection electrodes}

\begin{figure}
    \centering
    \includegraphics[width=.7\textwidth]{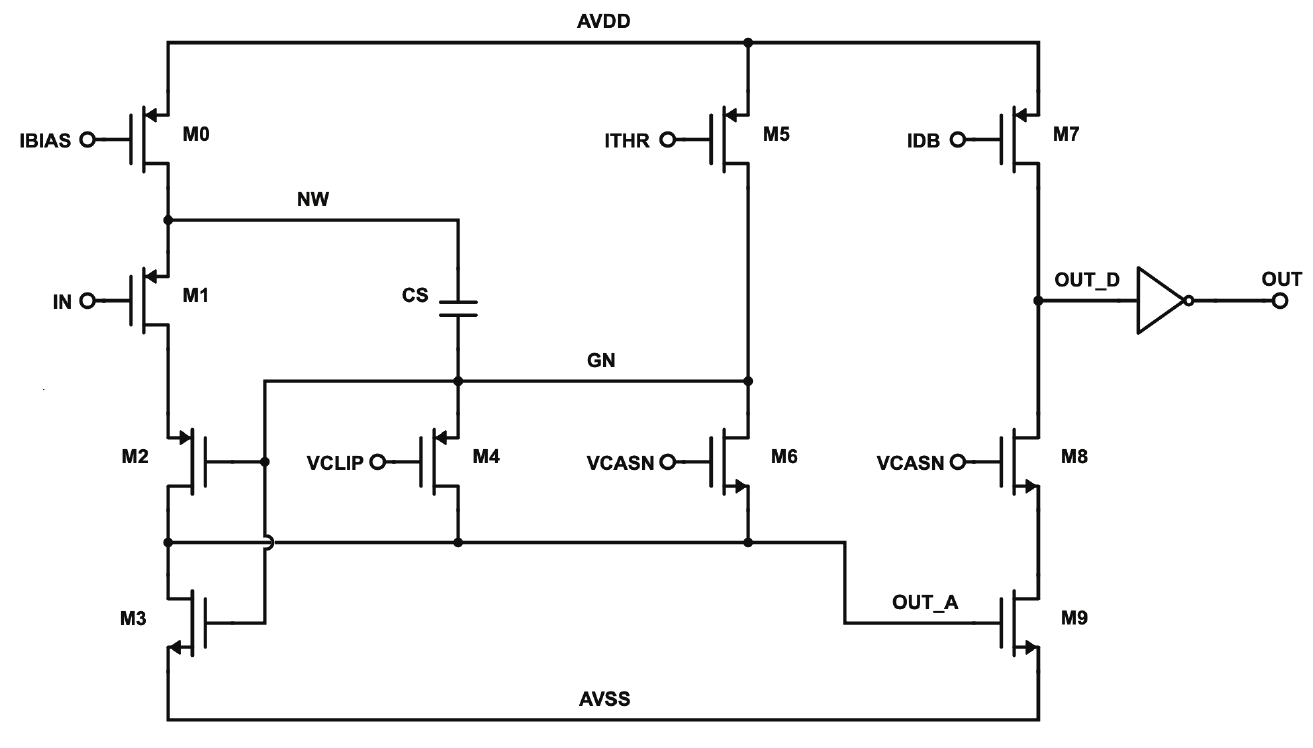}
     \caption{Analog front-end circuit of the MALTA and Mini-MALTA sensors.
    }
    \label{fig:circuit}
\end{figure}

The small collection electrode minimizes input capacitance to achieve a high Q/C ratio at the circuit input. In the case of a 25 $\mu$m thick sensitive layer we expect a most probable ionization charge of around 1500 e$^{-}$. 
To calculate the expected ionisation charge for our sensor, we assume an ionisation charge of 63 electron-hole pairs per $\mu$m path length~\cite{Meroli}.
With a total electrode capacitance of 5 fF this results in a voltage step of around 50 mV. This offers the possibility of using an open-loop voltage amplifier as the first amplification stage, instead of the conventional charge-sensitive amplifier scheme with a feedback capacitor, to save space and simplify the circuit. The collection electrode input voltage is reset after a particle hit using a diode-circuit or a PMOS-transistor. The front-end (FE) amplifier output connects to a discriminator, which produces the digital signal for a hit pixel. The discriminator threshold is set globally for the full sensor.  The analog FE circuit is shown in Figure~\ref{fig:circuit}. 


Initial measurements of this circuit on the MALTA sensor revealed significant Random-Telegraph-Signal (RTS) noise preventing lower threshold settings. This was attributed to the ``M3'' transistor being much smaller than on previous circuits. To verify this assumption the Mini-MALTA sensor includes sectors with the same ``M3'' size as on MALTA and larger. Measurements on the Mini-MALTA sensor confirmed that the larger transistor sizes significantly decreased the RTS noise, both before and after irradiation. In addition, the larger NMOS size also yielded a significantly larger gain and lower charge threshold for  the same settings. Further measurements confirmed the ``M3'' output conductance was higher than expected, which caused gain degradation for the FE. In addition it was found that for lower threshold settings, where the influence of this output conductance is larger, the spread on the gain and hence the threshold spread increased.

\subsection{Charge collection implant structures in Mini-MALTA}

\begin{figure}
    \centering
    \includegraphics[width=.39\textwidth]{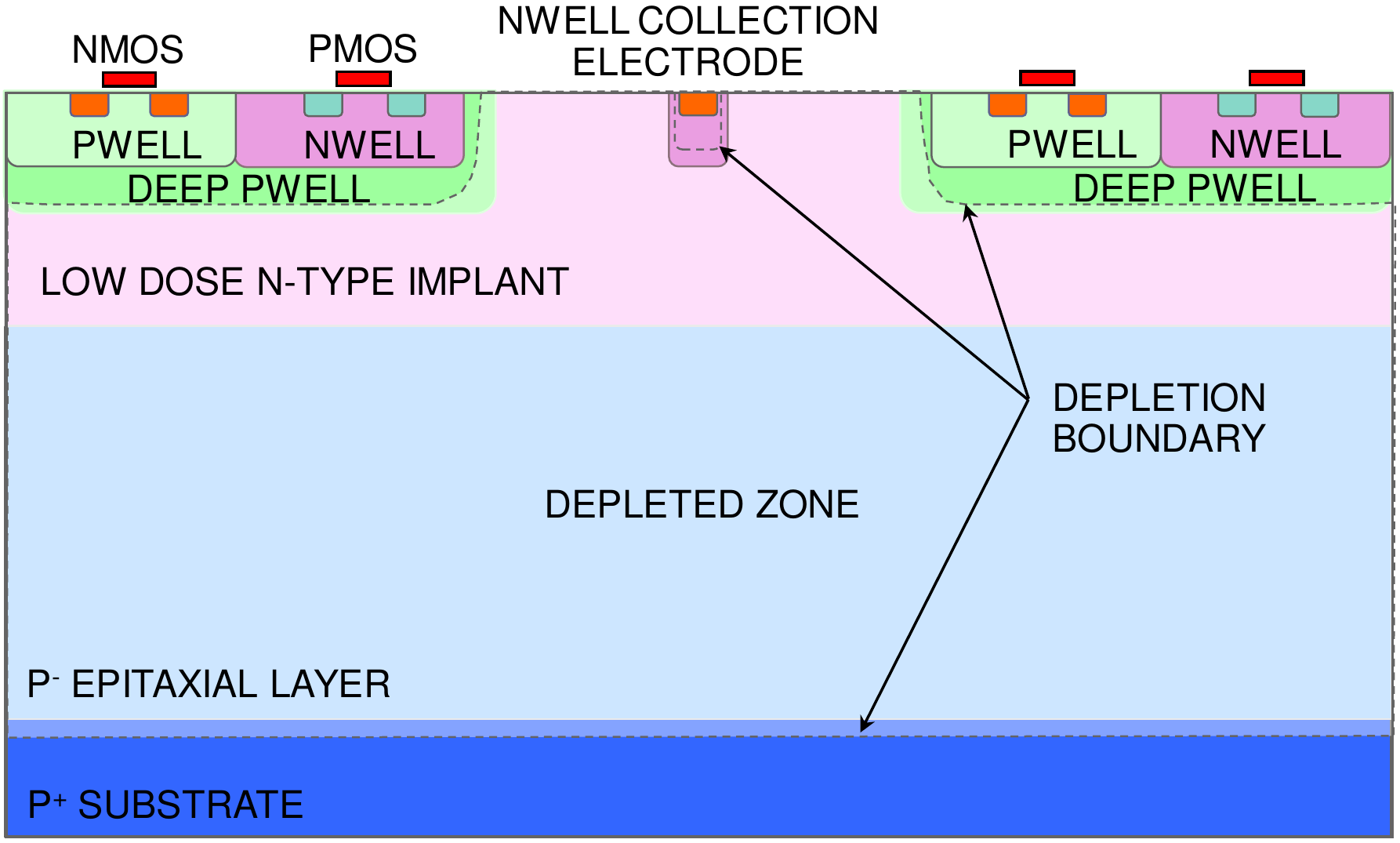}
    
    \includegraphics[width=.39\textwidth]{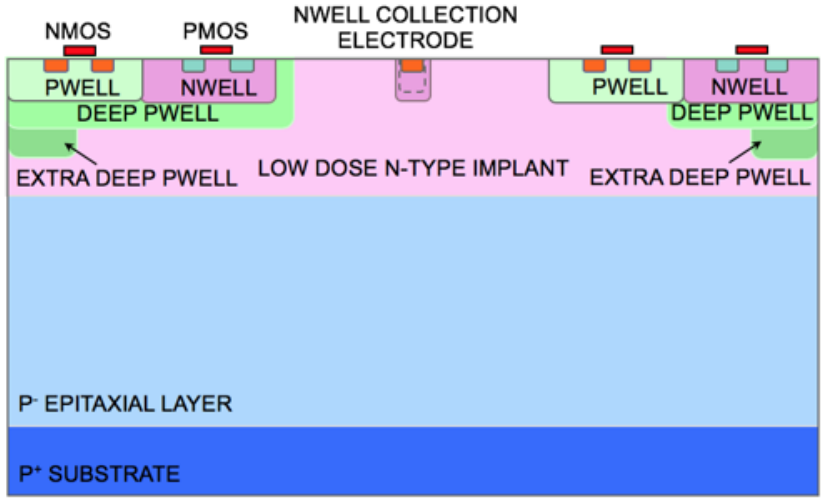}
    \includegraphics[width=.39\textwidth]{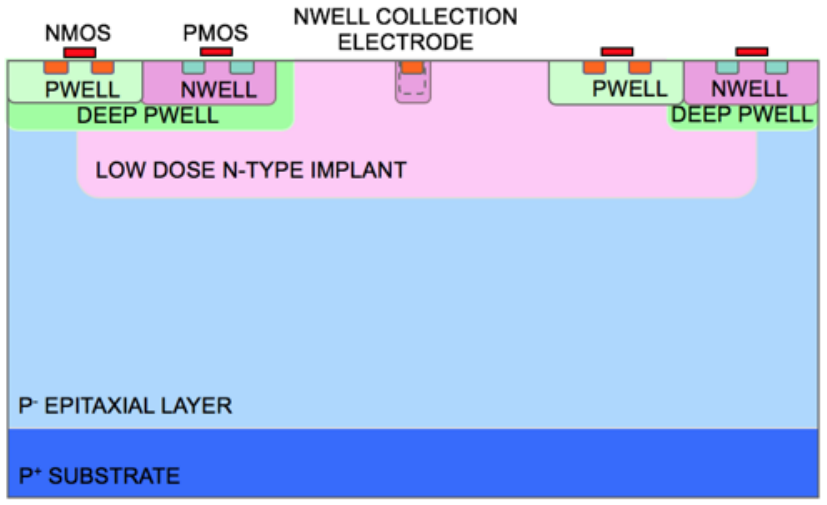}
    \caption{Cross section of the TowerJazz process: continuous n- layer (top), with extra deep p-well at the edge of the pixel (bottom left) and with the low dose n-implant removed (n$^-$ gap) at the edge of the pixel (bottom right). 
    }
    \label{fig:process_modification}
\end{figure}

Figure~\ref{fig:process_modification} (top) shows the cross-section of the TowerJazz process with a special low-dose $n$-type implant addition across the full pixel matrix~\cite{Snoeys:2017hjn}. This implant generates the junction to deplete the epitaxial silicon layer (25 $\mu$m thickness). This n$^{-}$ layer separates the deep p-well of the pixel circuit from the p-type substrate. The substrate is reverse biased (0 to $-$20 V) to fully deplete the epitaxial layer of the entire pixel. We refer to this implant configuration as ``standard continuous n- layer''. The deep p-well is biased in our measurements at $-$2 V. The n$^{-}$ layer is depleted from its junction to deep-p-well and p-type epitaxial layer. Depending on the choice of n$^{-}$ layer doping concentration the n$^{-}$ layer may not be fully depleted near the electrode at $-$2 V, which influences capacitance and gain. Changing the p-well voltage to $-$6 V in future prototypes is expected to improve charge collection and reduce capacitance.   

The TowerJazz process can be further modified (as shown in Figure~\ref{fig:process_modification} bottom row) by adding a gap in the low dose n-layer through a mask change (lower right) or adding an additional production process compatible deep p-type implant (lower left). We refer to these configurations as ``n$^{-}$ gap'' and ``extra deep p-well'' configurations, respectively. The purpose of these modifications is to improve the charge collection at the pixel edges and corners through the creation of a stronger lateral field, which focuses the ionization charge towards the collection electrode. The design of these implant structures has been optimized in TCAD simulations~\cite{Munker:2019vdo}, which indicate that these modifications significantly improve the charge collection at the pixel boundaries.


\subsection{The Mini-MALTA sensor design}
The Mini-MALTA chip matrix contains $16\times64$ square pixels with a pitch size of 36.4 $\mu$m. The full chip measures  $1.7 \times 5$ mm$^2$ including periphery blocks for data handling, sensor configuration and biasing. 

\begin{figure}
    \centering
    \includegraphics[width=.49\textwidth]{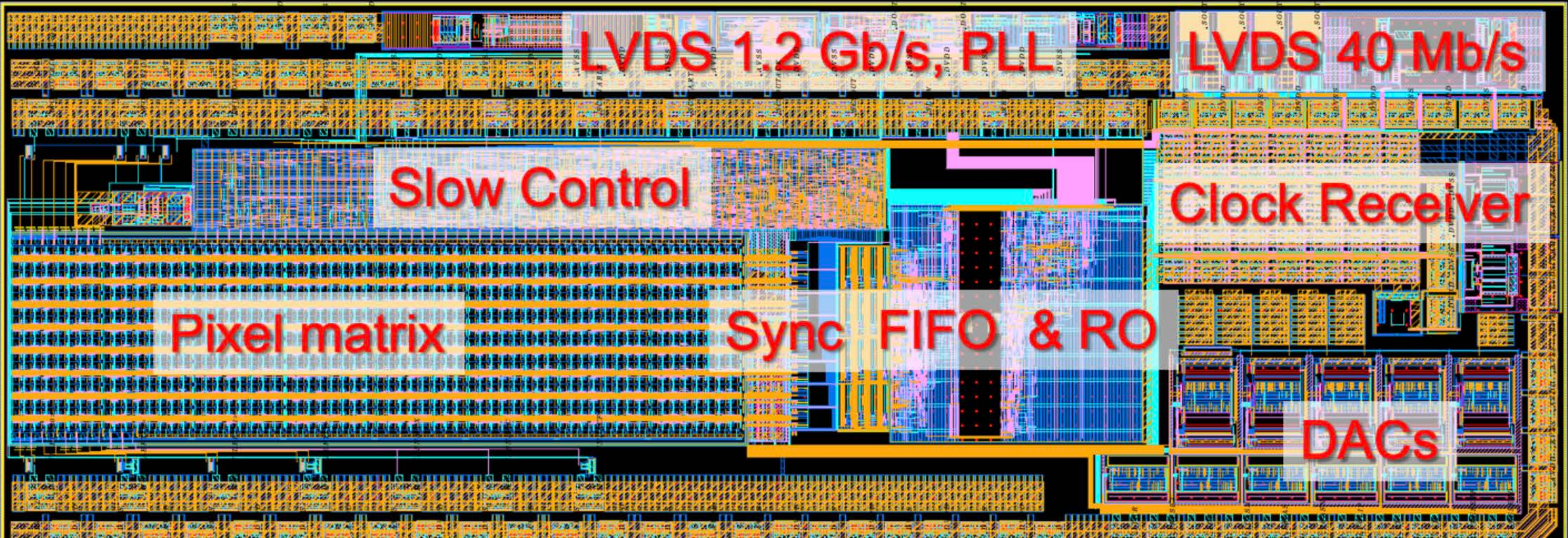}
    \includegraphics[width=.35\textwidth]{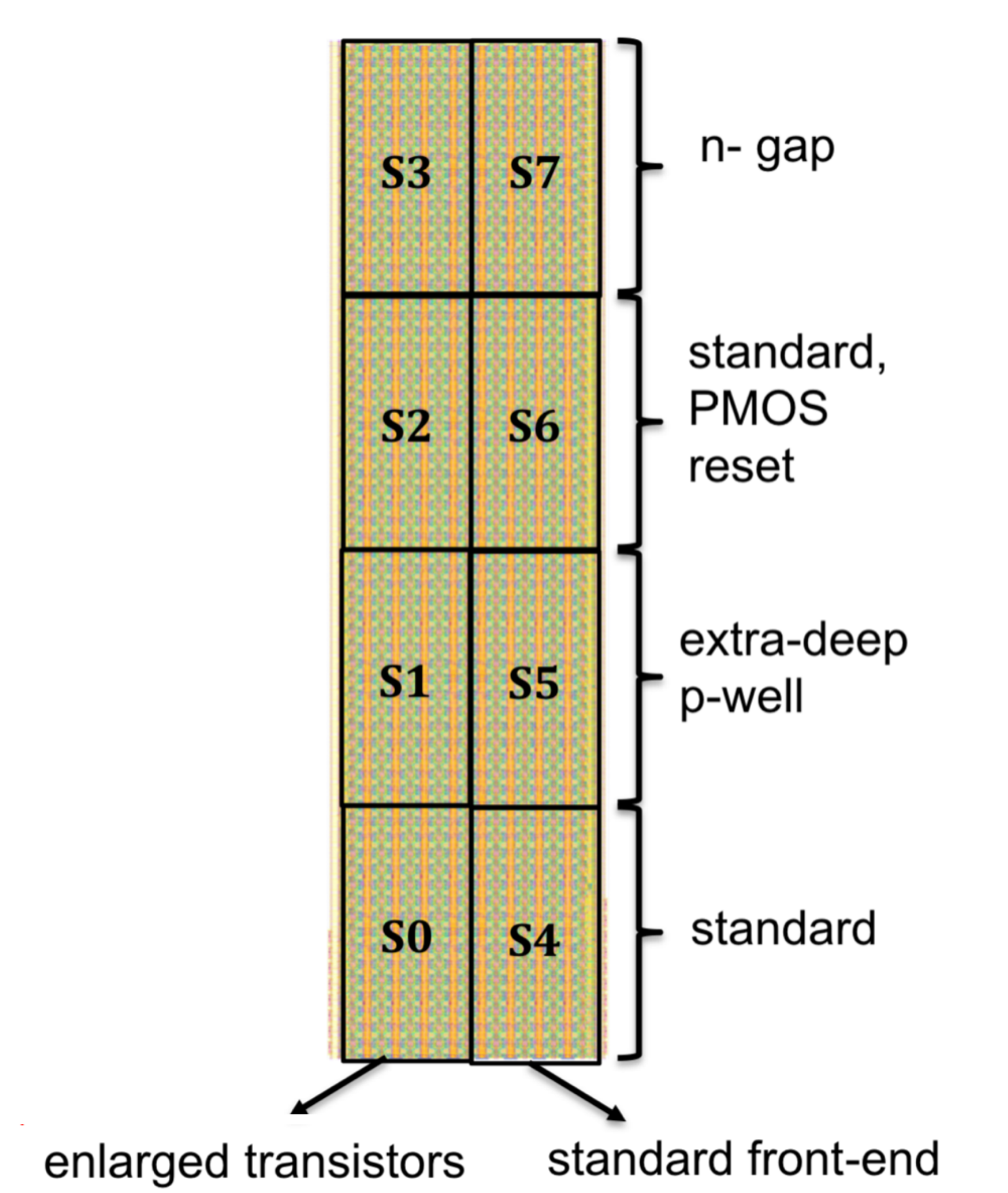}
    \caption{Top view of the full Mini-MALTA chip (left). Layout of the Mini-MALTA pixel matrix at the read-out level and pixel sub-groups of 8$\times$16 pixels (right).}
    \label{fig:minimalta}
\end{figure}

The Mini-MALTA sensor comprises eight different pixel flavours differing in analog FE design, reset mechanism and electrode/well geometries, as detailed in Table~\ref{tab:mini-malta-sectors}. They are implemented in the matrix as 8$\times$16 pixel groups as illustrated in Figure~\ref{fig:minimalta}. 

\begin{table}[h!]
\centering
\begin{tabular}{|c| c| c| c|} 
 \hline
 Sector ID & Pre-amp design & Reset type & Implant configuration \\ [0.5ex] 
 \hline\hline
 0 & enlarged transistor FE & diode reset & continuous n$^-$ layer \\ 
 1 & enlarged transistor FE & diode reset & extra deep p-well\\
 2 & enlarged transistor FE & PMOS reset & continuous n$^-$ layer \\
 3 & enlarged transistor FE & diode reset  & n$^-$ gap \\
 4 & standard transistor FE & diode reset & continuous n$^-$ layer \\ 
 5 & standard transistor FE & diode reset & extra deep p-well\\
 6 & standard transistor FE & PMOS reset & continuous n$^-$ layer \\
 7 & standard transistor FE & diode reset  & n$^-$ gap \\
  \hline  \hline
\end{tabular}
\caption{Pixel sub-groups in the Mini-MALTA pixel matrix.}
\label{tab:mini-malta-sectors}
\end{table}

Each pixel contains a charge collection electrode formed by an n-well of 3 micron diameter, contacted by an n+ region to the readout circuit. The n-well itself forms the contact to the lightly doped n-implant below. Reverse bias depletes the lightly doped n- region and also some part of the n-well. The undepleted remainder of the n-well then collects the signal charge.


The collection electrode is connected to the analog FE circuit, which is located in a separate deep p-well. The output of the analog circuit connects to the digital circuit, which buffers the discriminator output of the FE-circuit and transmits a digital signal to the end-of-column logic. The pixels are organized in groups of $2 \times 8$. 
Hits from pixels are sent to dedicated logic, common within the group, generating a reference pulse (1 bit) as well as a pixel address (16 bit where each bit corresponding to a particular pixel out of the 16 pixels in each group) and group address (5 bit) signal on the 22-bit wide double-column bus. Digital signals on this bus are 2ns wide and are transmitted asynchronously to the periphery~\cite{Cardella:2019ksc}. No clock is distributed over the matrix to minimize power consumption and avoid cross-talk between digital and analog circuit. The digital signal from the pixel is stored in the end-of-column logic memory: The data from the groups are stored asynchronously into 16 synchronization memories, which are then read out synchronously with the external 320 MHz clock using a priority encoder.
Inside the synchronization memory, the precise time-of-arrival information is added by latching the value of a 4-bit counter running at 640~MHz.
When reading out the hits, the priority encoder gives priority to leftmost pixel groups (see Figure~\ref{fig:minimalta} right) in case of simultaneous hits in multiple memories. In the end-of-column logic the 4-bit double-column address is added to the data words, along with 3 bit bunch-crossing counter information (BCID) which is generated in the synchronization memories by using the external 40 MHz clock. The data from the end-of-column logic is stored into a FIFO with a depth of 64 48-bit words.

By default, the Mini-MALTA operates in 'fast' readout mode where the 48-bit 8b/10b encoded data is sent off from the chip at 1.2 Gbps. In this case a 600 MHz clock, which is generated from the external 40 MHz clock,  is used to send the data at double data rate. The Mini-MALTA sensor also offers the possibility to output the data at 40 Mbps for setups where fast signal transmission is not necessary nor desirable. This mode was used for beam tests reported here. To mark the beginning and end of the sent 48-bit words, a dedicated 'acknowledge' output signal is used, which is active only during the $48\times25$ ns when the data is being transmitted. 

\section{Laboratory measurements}

\subsection{Signal response using $^{55}$Fe source}
The response of the sensors is verified using photons from a $^{55}$Fe source and dedicated ``test-pixels'' with analog readout.
$^{55}$Fe produces photons with two characteristic lines, K$_{\alpha}$ and K$_{\beta}$, having energies of 5.9 keV and 6.5 keV respectively. The K$_{\alpha}$ peak is the dominant decay mode. The Mini-MALTA test-pixels allow to probe the analog signal before the discriminator (``OUT-A'' in Figure~\ref{fig:circuit}) and at the collection electrode (``IN'' in Figure~\ref{fig:circuit}).

\begin{figure}[!htb]
    \includegraphics[width=.505\textwidth]{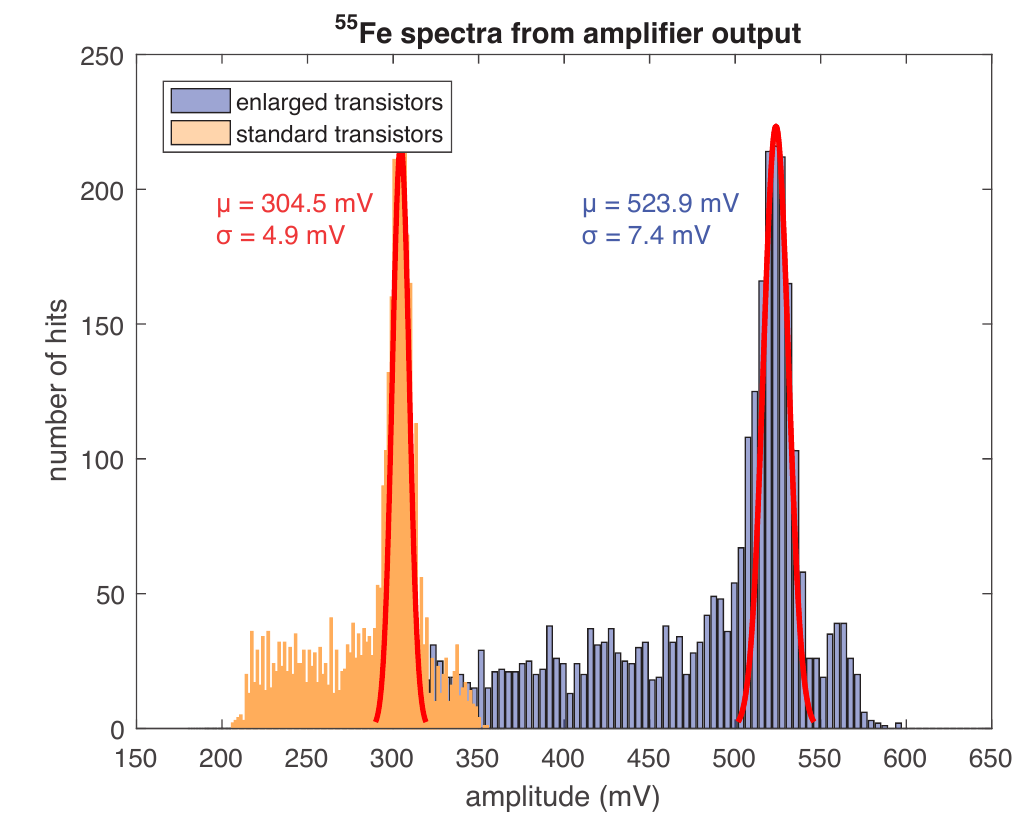}
    \includegraphics[width=.54\textwidth]{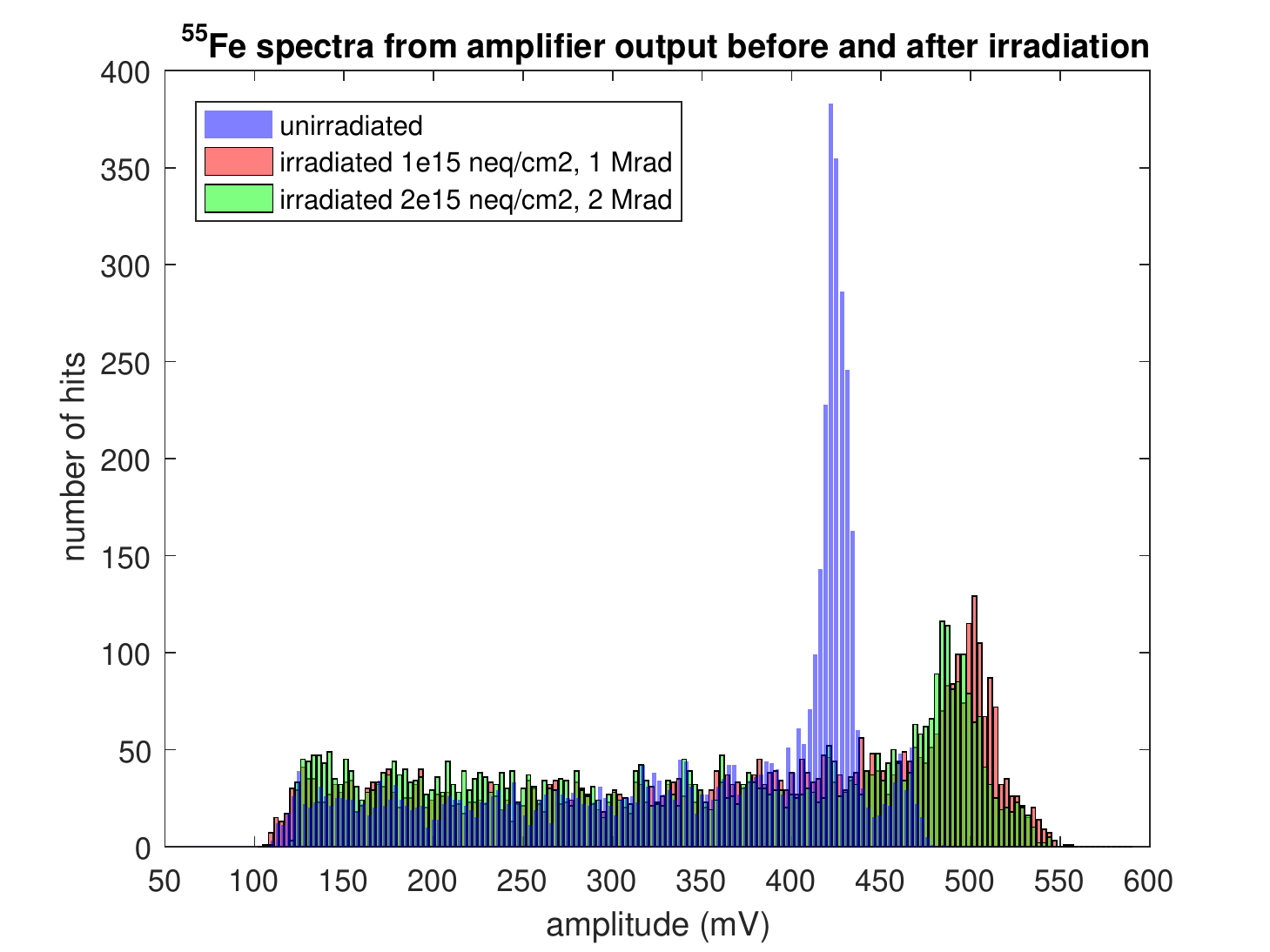}
    \caption{Signal amplitude distribution in response to an $^{55}$Fe source as measured on monitoring pixels after the amplifier. The left plot compares the response of the unirradiated sensor for enlarged (sector 0) and standard transistor FE (sector 5). The right plot shows the response for the unirradiated and neutron-irradiated Mini-MALTA samples at sector 0.
    The chips were operated at $-6$ V SUB voltage at $-20^{\circ}$C temperature and with identical FE settings for each plot.}
    \label{fig:fe-55}
\end{figure}

Figure~\ref{fig:fe-55} (left) shows the signal amplitude spectrum from the test pixels measured on ``OUT-A'' with enlarged NMOS transistors for unirradiated Mini-MALTA samples in comparison to preamplifiers with standard (minimal-size) transistors. 
As explained earlier, enlarging the NMOS transistor ``M3'' significantly improves gain and gain uniformity and also RTS noise and therefore allows operation at lower thresholds.
Clear peaks from K$_{\alpha}$ and K$_{\beta}$ lines of $^{55}$Fe are visible for the unirradiated chip.

Figure~\ref{fig:fe-55} (right) shows the signal amplitude spectrum from the test pixels measured on ``OUT-A'' with enlarged NMOS transistors for unirradiated and neutron irradiated sensors. The sensors received for $1\times10^{15}$ and 2$\times10^{15}$ 1 MeV n$_{eq}$/cm$^{2}$ fluence an additional 1 Mrad and 2 Mrad of TID through the gamma background in the reactor. For the irradiated Mini-MALTA sensors, a clear increase in noise is observed, resulting in an increased width of both peaks.
Moreover, there is a shift in the mean peak position visible after irradiation, which suggests that the irradiated Mini-MALTA sensors increase in gain by 20\% when operated at the same preamplifier setting. For comparison the amplitude spectrum was also measured at the collection electrode, so at the input of the readout circuit (``IN''), for unirradiated and irradiated sensors, which also showed a 20\% higher $^{55}$Fe-signal for the irradiated sensors. 
Most likely it points to a change in sensor capacitance due to changes in effective doping concentration, rather than a gain change in the readout circuit.

 The chip configurations used in Figure~\ref{fig:fe-55} (left) and in Figure~\ref{fig:fe-55} (right) were slightly different, resulting in some shift in mean peak position for unirradiated sensors and regions with enlarged NMOS transistors.

\subsection {Threshold measurement and tuning}

The threshold in the full pixel matrix is modified through dedicated on-chip 8-bit DACs which adjust the biasing of the pre-amplifier stage and in-pixel discriminator. The main threshold adjustment is carried out through the discriminator bias current (``IDB''), which is set globally for the full matrix.
For the beam test measurements described in the following sections, the chips are tuned to different thresholds to test efficiency as function of applied threshold.

The resulting threshold is measured for each pixel by injecting pulses with varying voltage amplitude on an in-pixel test capacitance (0.23~fF) at the input of each preamplifier. For each pixel, 200 injections per test pulse amplitude are performed, and the threshold point corresponding to 50\% occupancy (50\% probability of a pixel being fired) is determined by fitting a Gauss error function to the hit occupancy S-curve. 
The width parameter (to deduce the value of equivalent noise charge ENC) is also extracted from the fit.

\begin{figure}
    \centering
    \includegraphics[width=.49\textwidth]{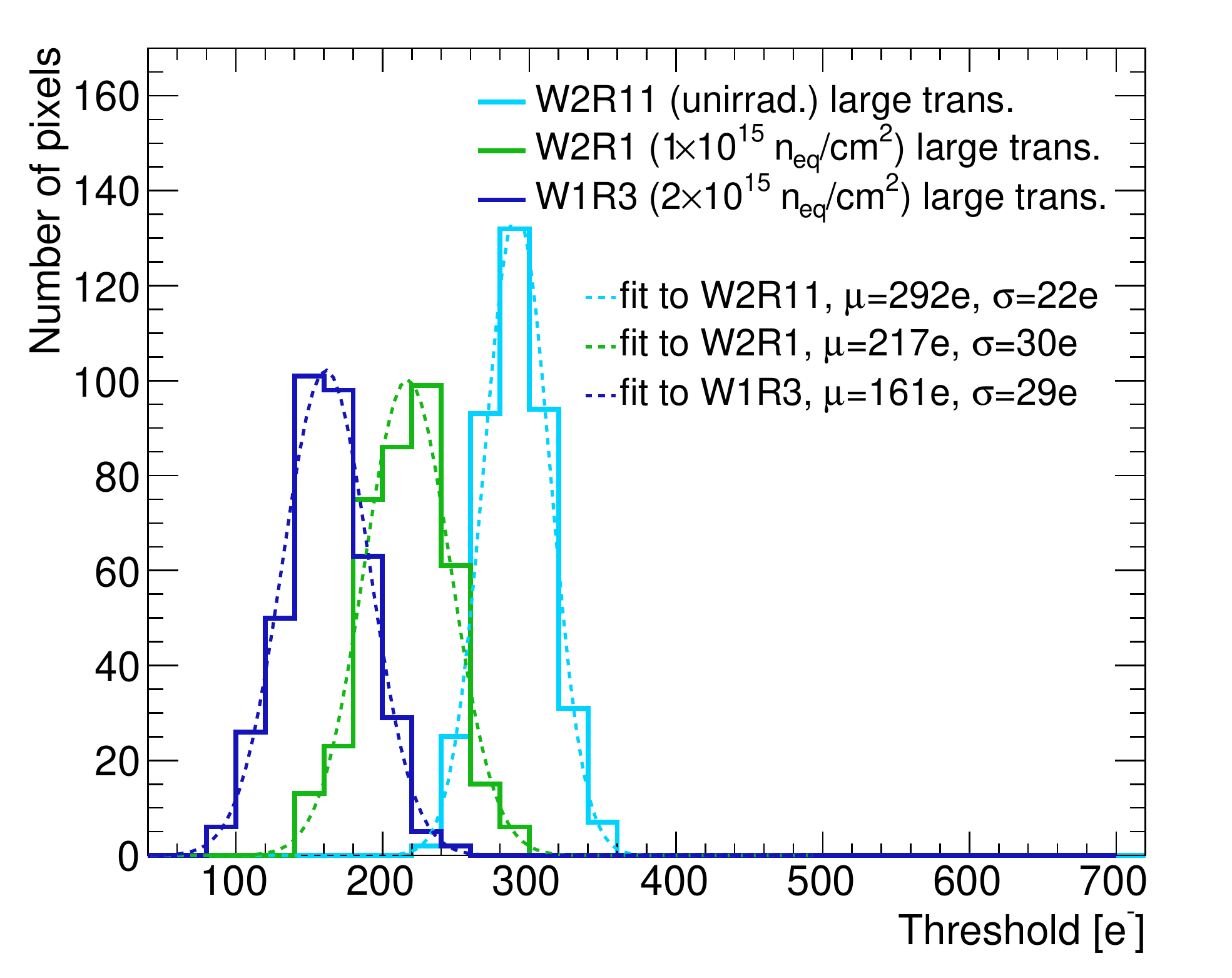}
    \includegraphics[width=.49\textwidth]{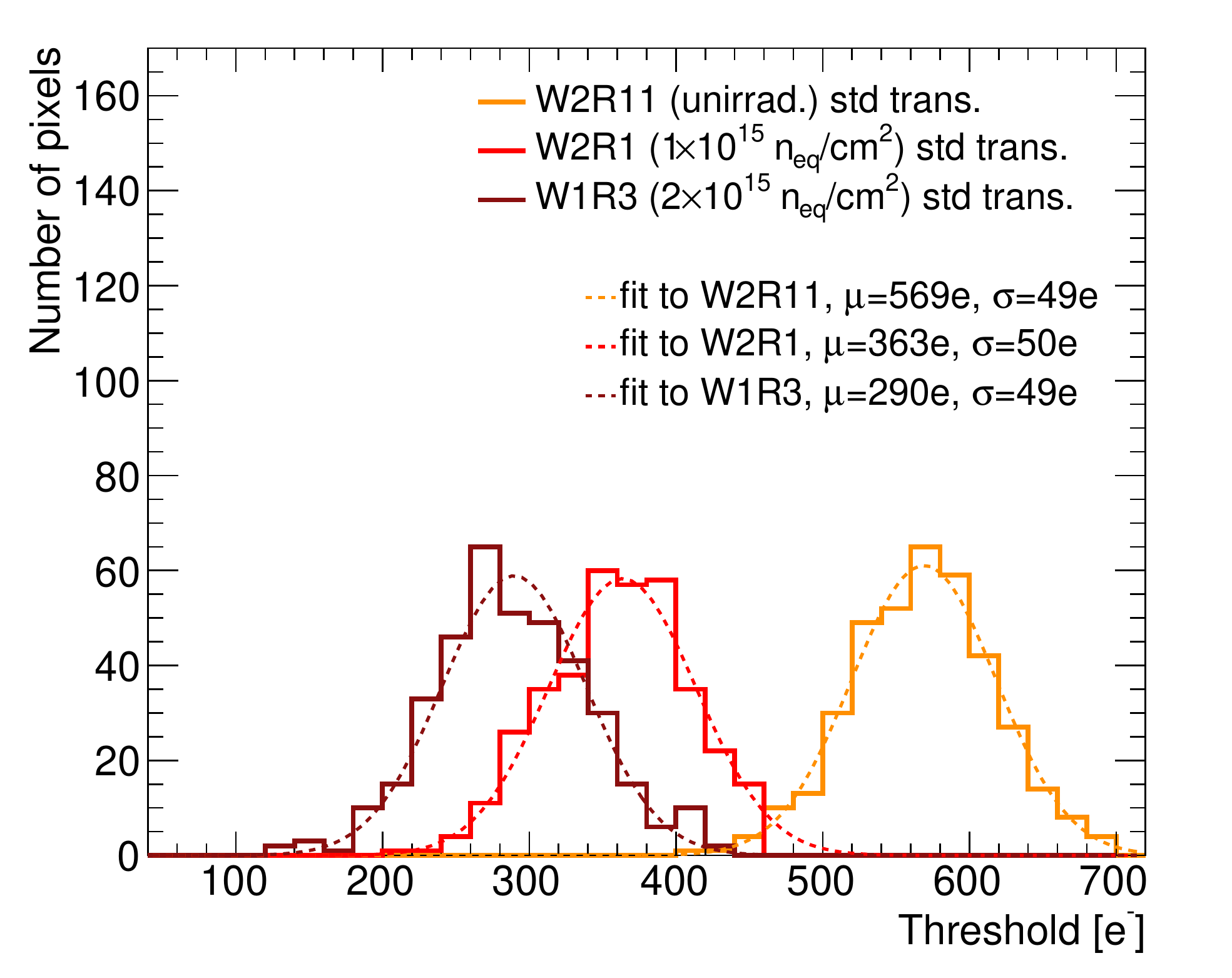}
    \caption{Threshold distributions for unirradiated and neutron-irradiated Mini-MALTA samples at 1$\times$10$^{15}$ and 2$\times$10$^{15}$ 1 MeV n$_{eq}$/cm$^{2}$. 
    Gaussian fits to threshold distributions are shown as dashed lines.
    The default chip tuning configuration is used (IDB=100) for unirradiated and irradiated sensors in the plots. Sensor regions with enlarged (left) and standard (right) transistors are shown. }
    \label{fig:threshold_comparison}
\end{figure}

Figure~\ref{fig:threshold_comparison} shows the measured threshold distributions for unirradiated and neutron irradiated Mini-MALTA samples, separately for sensor regions with standard (right plot) and enlarged transistors (left plot). Default chip tuning configuration is used for all sensors. The sensors were operated at $-$6 V substrate voltage in a climatic chamber set at $-20^{\circ}$C.
The gaussian fits are performed to the threshold distributions for each chip.
For regions with standard transistors the average threshold extracted from fit decreases from 570e$^{-}$ (unirradiated) to 360e$^{-}$ ($1\times10^{15}$ 1 MeV n$_{eq}$/cm$^{2}$) and 290e$^{-}$ ($2\times10^{15}$ 1 MeV n$_{eq}$/cm$^{2}$). As already mentioned most of this change is due to a change in the sensor capacitance rather than in the readout circuit. 
In all regions with enlarged transistors the average threshold values are systematically lower due to the higher gain of the FE circuit with enlarged transistors: 290e$^{-}$ (unirradiated), 220e$^{-}$ ($1\times10^{15}$ 1 MeV n$_{eq}$/cm$^{2}$) and 160e$^{-}$ ($2\times10^{15}$ 1 MeV n$_{eq}$/cm$^{2}$). 
The threshold dispersion extracted from the gaussian fit is larger for regions with standard transistors (around 50e$^{-}$) than for regions with enlarged transistors (20-30e$^{-}$). 

\begin{figure}[!htb]
    \centering
    \includegraphics[width=.49\textwidth]{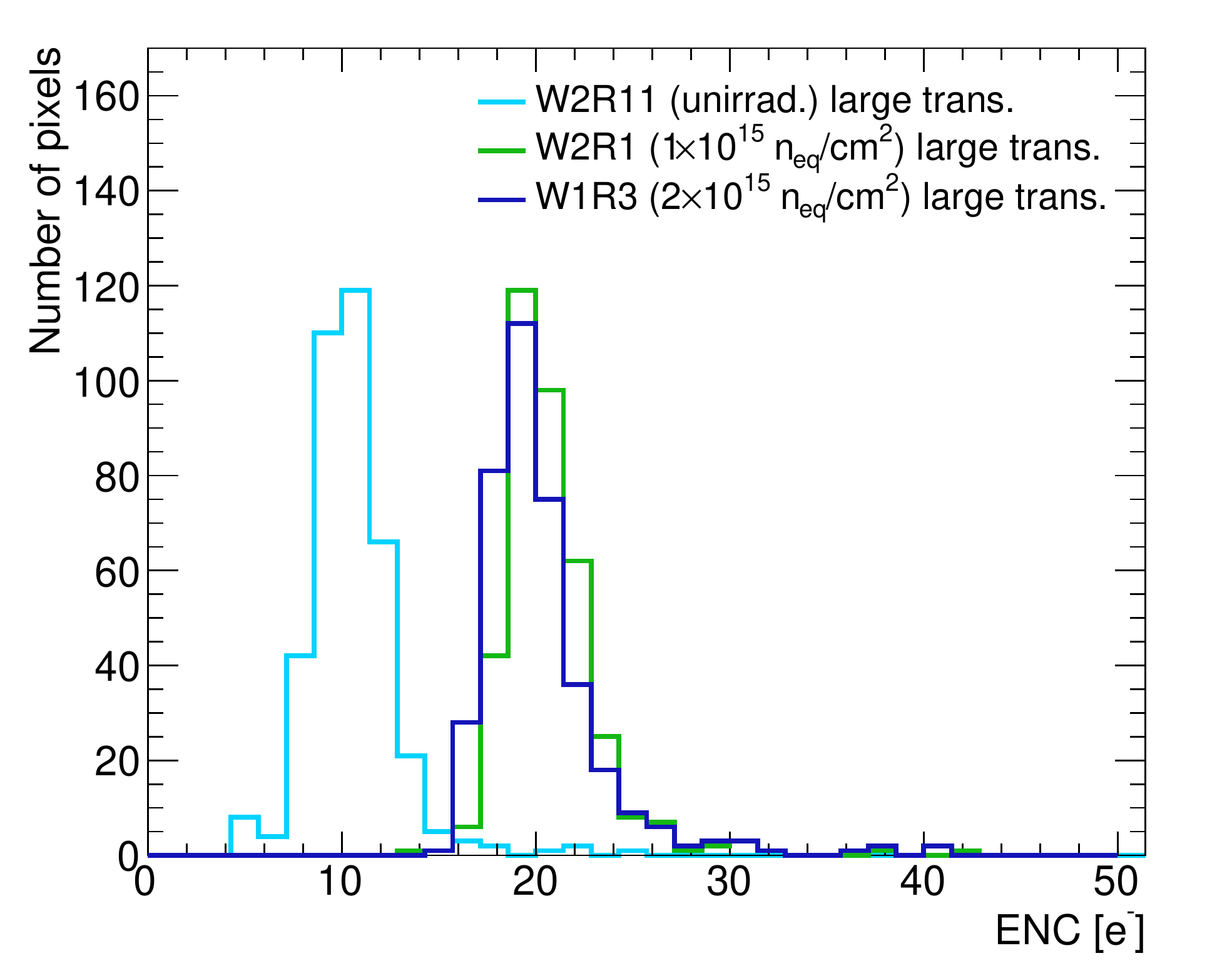}
    \includegraphics[width=.49\textwidth]{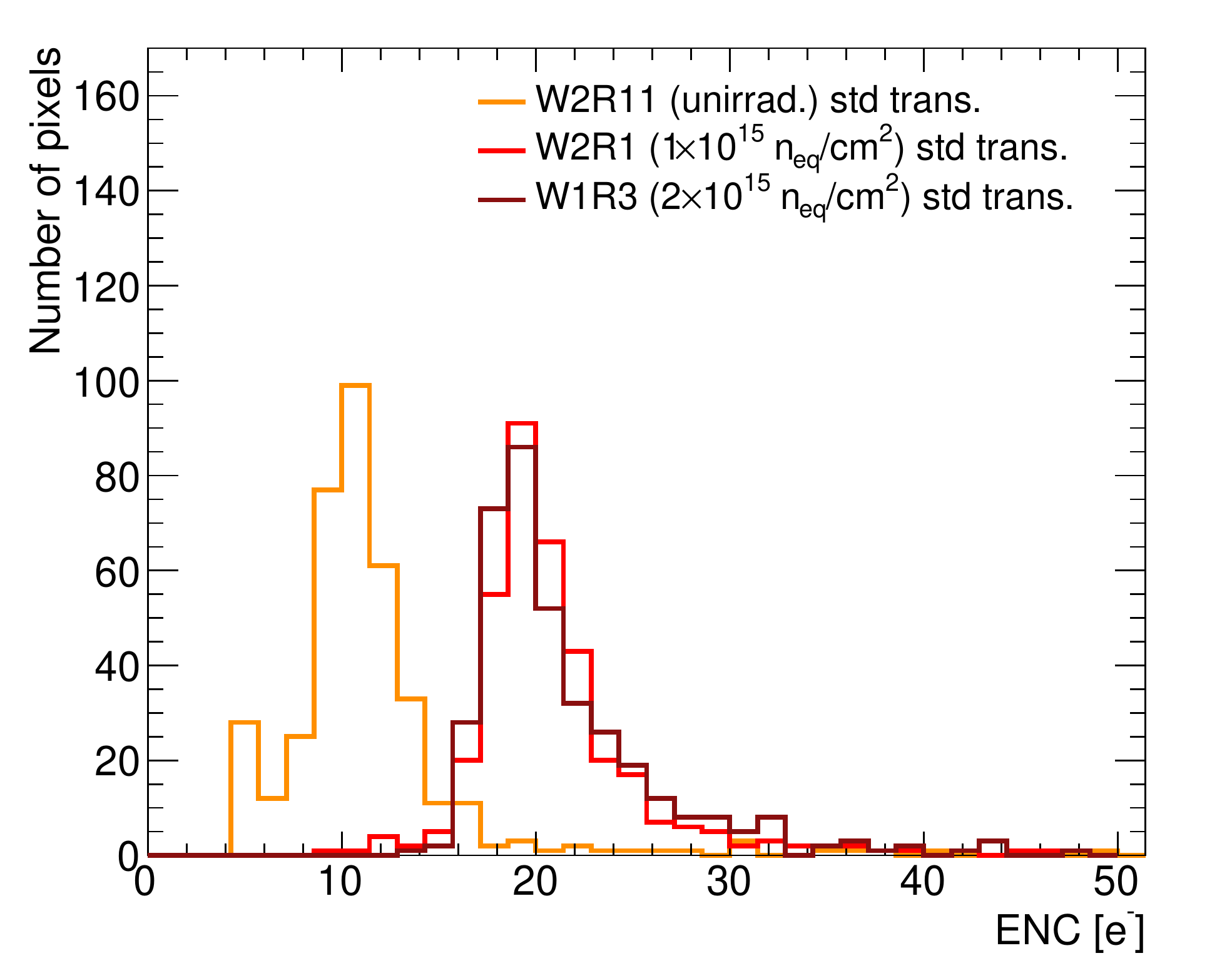}
    \caption{ENC distributions for unirradiated and neutron-irradiated Mini-MALTA samples at 1$\times$10$^{15}$ and 2$\times$10$^{15}$ 1 MeV n$_{eq}$/cm$^{2}$. Sensor regions with enlarged (left) and standard (right) transistors are shown. The chips were operated at $-6$ V SUB voltage, $-20^{\circ}$C and with same settings.}
    \label{fig:enc_comparison}
\end{figure}

The ENC distributions for unirradiated and irradiated Mini-MALTA sensors are shown in Figure~\ref{fig:enc_comparison}.
The most probable ENC value before irradiation is 11e$^{-}$ and increases to $\approx$20e$^{-}$ for chips after irradiation. 
Moreover, the ENC dispersion is lower in the regions with enlarged transistors ($\sigma_{\textrm{ENC}}\approx 3$e$^{-}$). The ENC distributions in sectors with standard transistors ($\sigma_{\textrm{ENC}}\approx 5$e$^{-}$) show a significant tail of pixels with high noise. We attribute this to Random Telegraph Signal noise (RTS) due to the use of minimal-size transistors in parts of the analog circuit. The measurements on sectors with enlarged transistors show that increasing the size of these transistors significantly reduces RTS noise and ENC dispersion as illustrated in Figure~\ref{fig:enc_comparison} (left).

\subsection{Noise occupancy}
\begin{figure}[!htb]
    \centering
    \includegraphics[width=.49\textwidth]{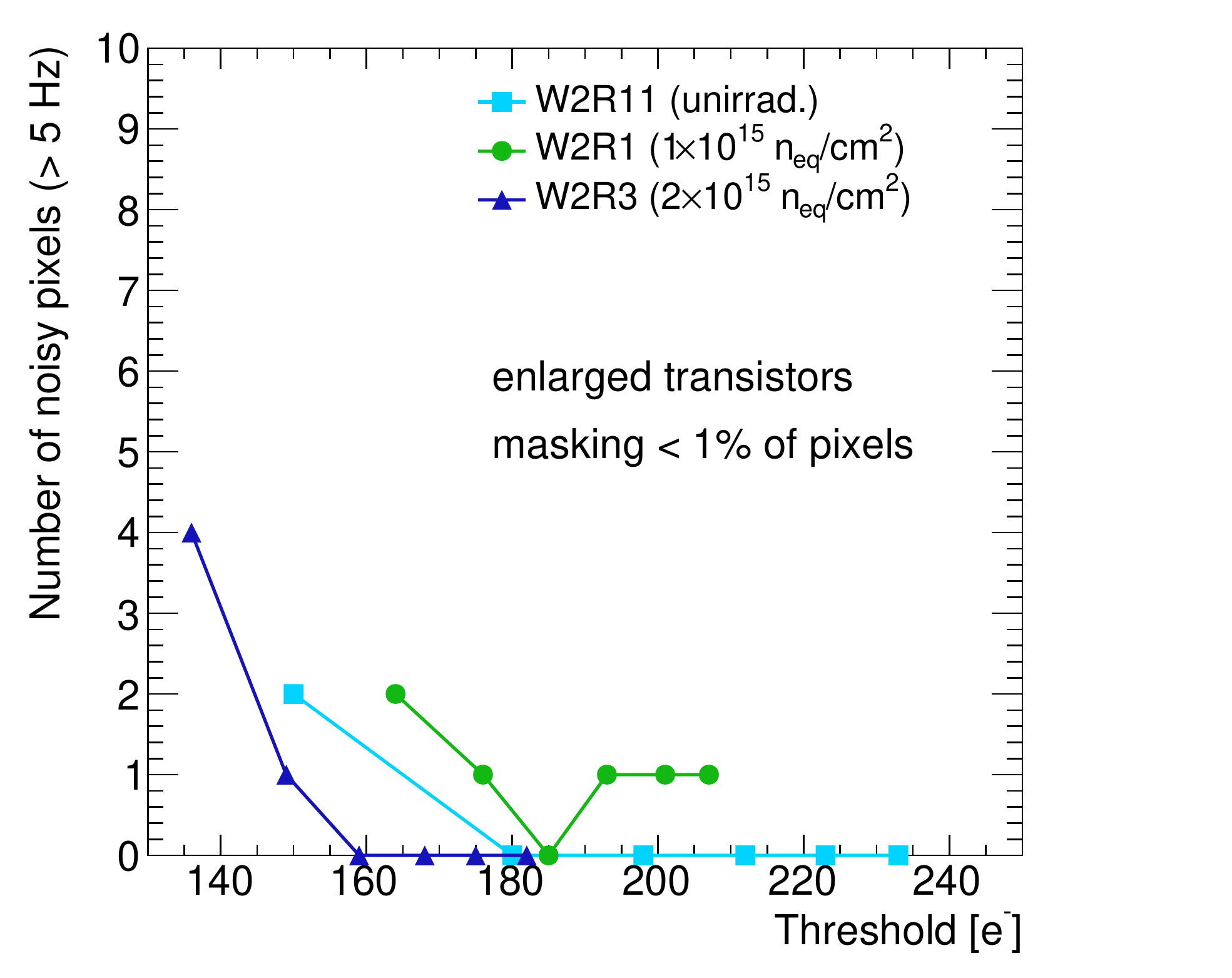}
    \includegraphics[width=.49\textwidth]{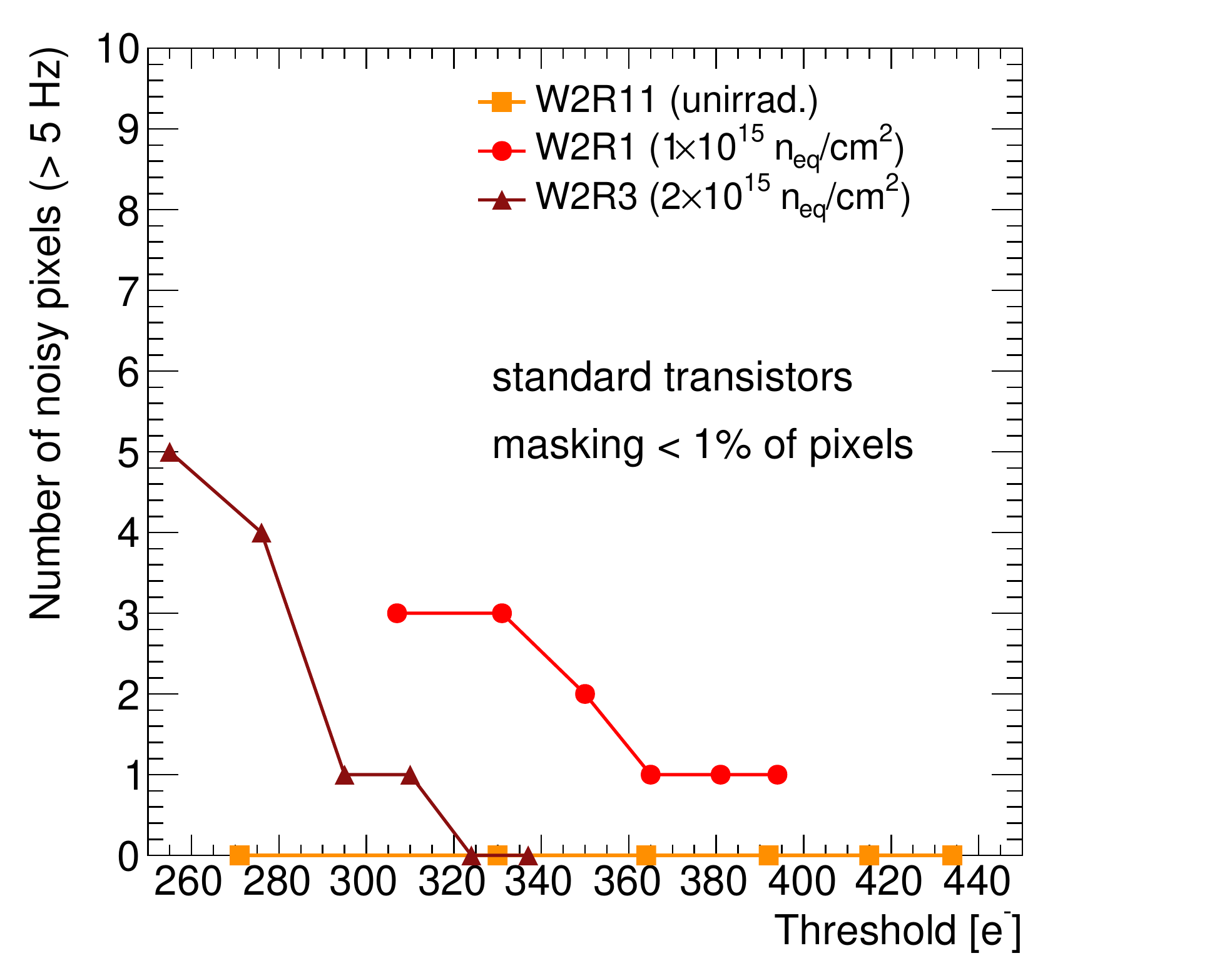}
    \caption{Distribution of the number of noisy pixels (with rate above 5 Hz) for unirradiated and neutron-irradiated Mini-MALTA samples at 1$\times$10$^{15}$ and 2$\times$10$^{15}$ 1 MeV n$_{eq}$/cm$^{2}$. Sensor regions with enlarged (left) and standard (right) transistors are shown. 
    For all chips, less than 1\% of total pixels in a given region are masked due to noise above 0.5 kHz, and these pixels are not counted in the procedure.
    The chips were operated at $-6$ V SUB voltage and $-20^{\circ}$C.}
    \label{fig:noise_comparison}
\end{figure}

The noise level for each Mini-MALTA chip is characterized by measuring the number of noisy pixels as a function of charge threshold. 
The noisy pixels with relatively large noise rate (above 0.5 kHz) are masked and are not counted in the following.
For each chip these masked pixels constitute less than 1\% of the total number of pixels.
Figure~\ref{fig:noise_comparison} presents the distribution of the number of noisy pixels as a function of charge threshold for various Mini-MALTA samples, separately for sensor regions with enlarged and standard transistors.
Typically for unirradiated Mini-MALTA chips, no more than 1 or 2 noisy pixels are observed even at very low thresholds ($\approx$100e$^{-}$).
For the neutron-irradiated Mini-MALTA sensors, the number of noisy pixels is low at higher thresholds (less or equal one), and the number grows with decreasing the threshold. 
At lowest thresholds there are 3--5 noisy pixels (out of 384 measured pixels) for each region of the neutron-irradiated Mini-MALTA sensors.

\section{Test beam setup and data analysis}

\subsection{Beam telescope arrangement}

The data presented in this study were recorded at the ELSA test beam facility at the University of Bonn. The ELSA synchrotron circulates one electron bunch with a maximum energy of about 3.5 GeV. 
The test beam is generated via a twofold conversion and the detectors were probed with 2 GeV electrons.

\begin{figure}
    \centering
    \includegraphics[width=.49\textwidth]{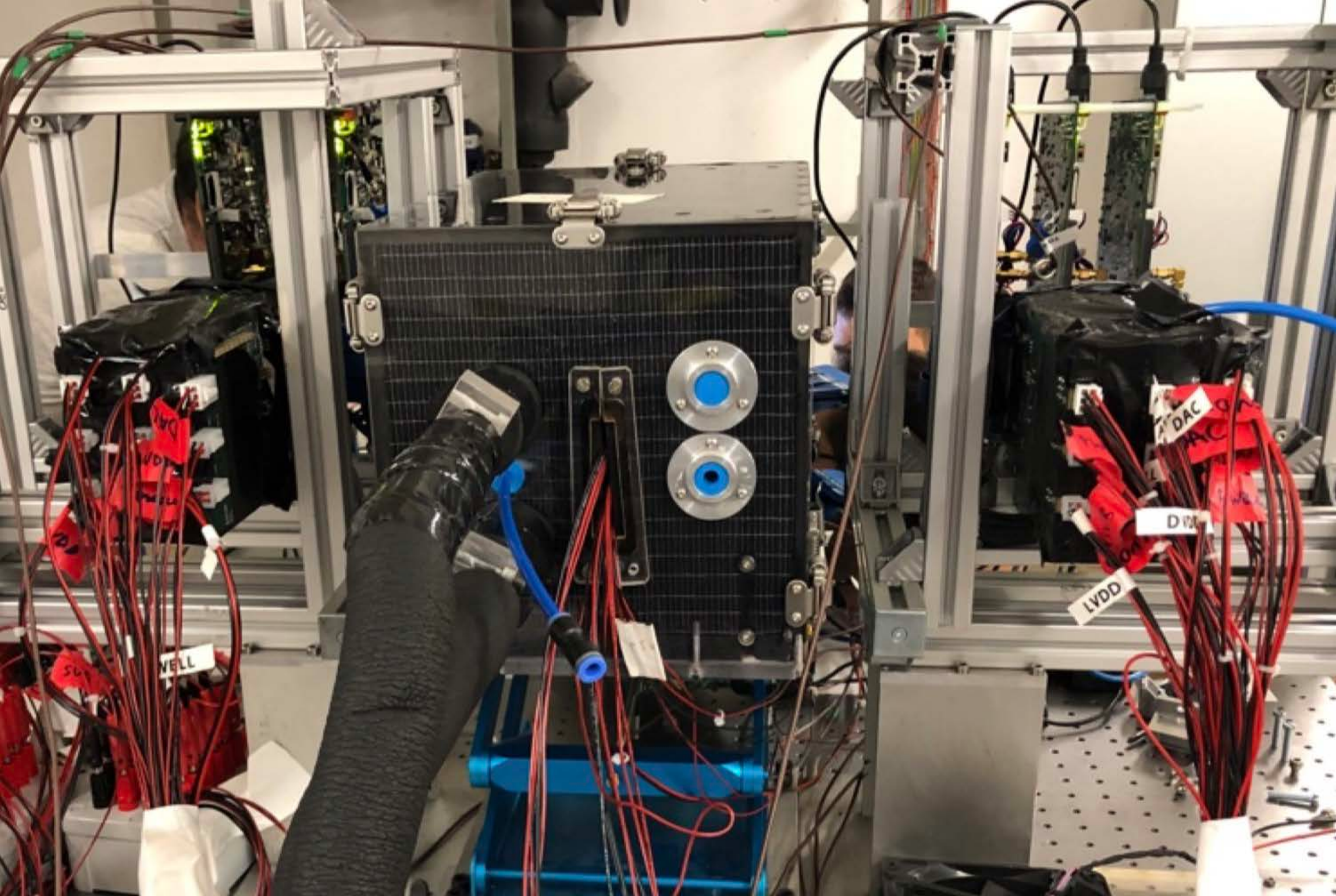}
    \includegraphics[width=.69\textwidth]{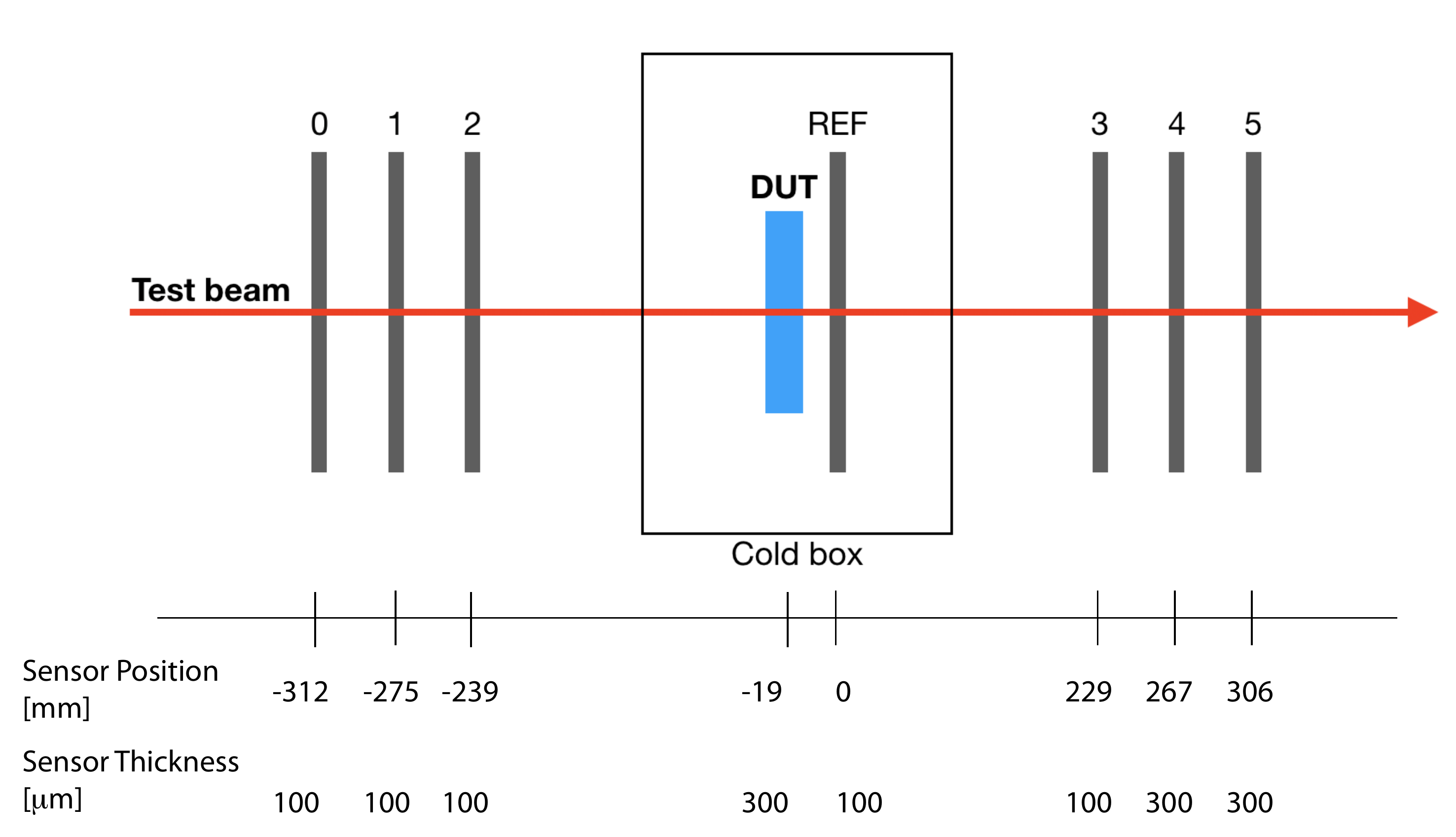}
    \caption{Photo of the test beam setup (top) and its graphical diagram (bottom) showing the beam telescope with seven MALTA tracking planes and the device under test (DUT) between the upstream (3 planes) , reference (1 plane, REF) and downstream (3 planes) arms of the telescope. The DUT and the reference MALTA plane are placed in the cold box, operated at $-20^{\circ}$C.}
    \label{fig:tb_setup}
\end{figure}

The test beam setup is shown in Figure~\ref{fig:tb_setup}. The beam telescope based on six MALTA planes was used.
Electrons first passed through 3 MALTA silicon detector planes, before entering the Device Under Test (DUT) and one extra MALTA reference plane (REF) placed close in front of the DUT. 
A REF plane close to the DUT improves the position resolution of the telescope when using the low-energy electron beam.
The DUT and REF planes are placed in the cold box which was operated at $-20^{\circ}$C.
Another 3 MALTA planes are placed downstream of the DUT and are used together with the upstream planes as reference system for track reconstruction. To minimize multiple scattering 100 $\mu$m thin sensors were used when possible.  Figure~\ref{fig:tb_setup} also gives the sensor position along the beam axis as well as the sensor thickness for each plane.

Each MALTA sensor provides a fast trigger signal when there is a hit on the plane. Upon coincidence of 2 or 3 planes, a trigger signal initiates the readout of the entire system. 
The data acquisition is performed using a custom read-out system based on the Xilinx VC707 boards and fast software~\cite{xilinx, Larrea_2015, Elmsheuser:2017xaw}.

\subsection {Track reconstruction and alignment}

Tracks are selected by requiring a hit on the third plane of the telescope in front of the DUT (plane '2' from Figure~\ref{fig:tb_setup}), and hits in the first two planes after the DUT (plane 'REF' and plane '3' from Figure~\ref{fig:tb_setup}).
Adjacent pixel hits are combined into clusters.
The track reconstructed from these three telescope layers is extrapolated to the plane of the DUT, taking into account multiple scattering by using the General Broken Lines (GBL) formalism~\cite{ kiehn_moritz_2019_2586736,Kleinwort:2012np}.
The track trajectory calculation uses the material description of the DUT and all telescope planes as well as the electron beam energy for multiple scattering estimation.

The alignment of telescope planes uses a two-step method. 
In the first step, a coarse alignment is performed, where the hits in X and Y of all neighboring telescope plane combinations are correlated and the resulting residual distributions are shifted with their means towards zero. 
The second step uses full telescope tracks and a $\chi^2$ minimization method for fine alignment. This returns the alignment parameters and uncertainties for each of the telescope planes. 

\begin{figure}
    \centering
    \includegraphics[width=.49\textwidth]{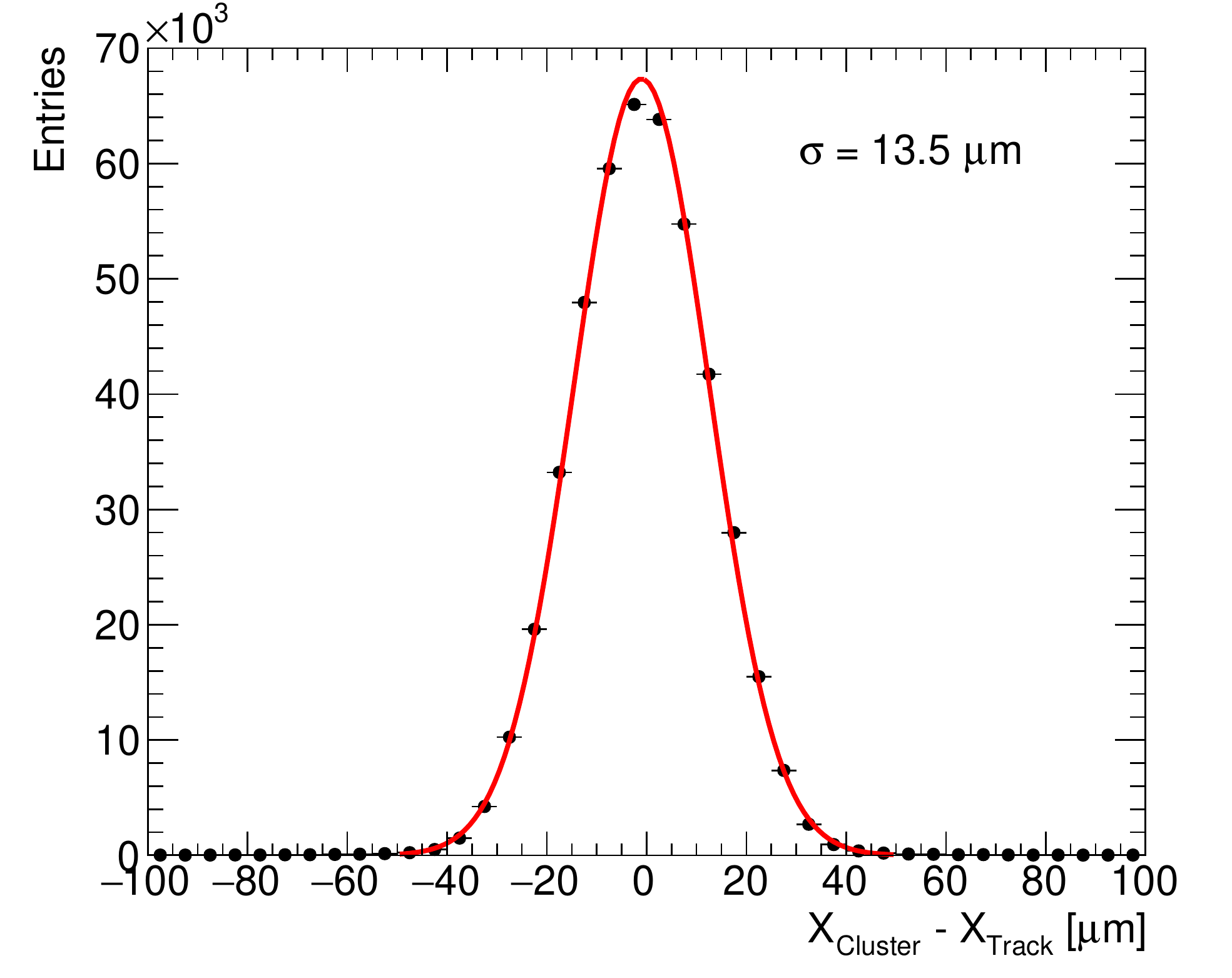}
    \includegraphics[width=.49\textwidth]{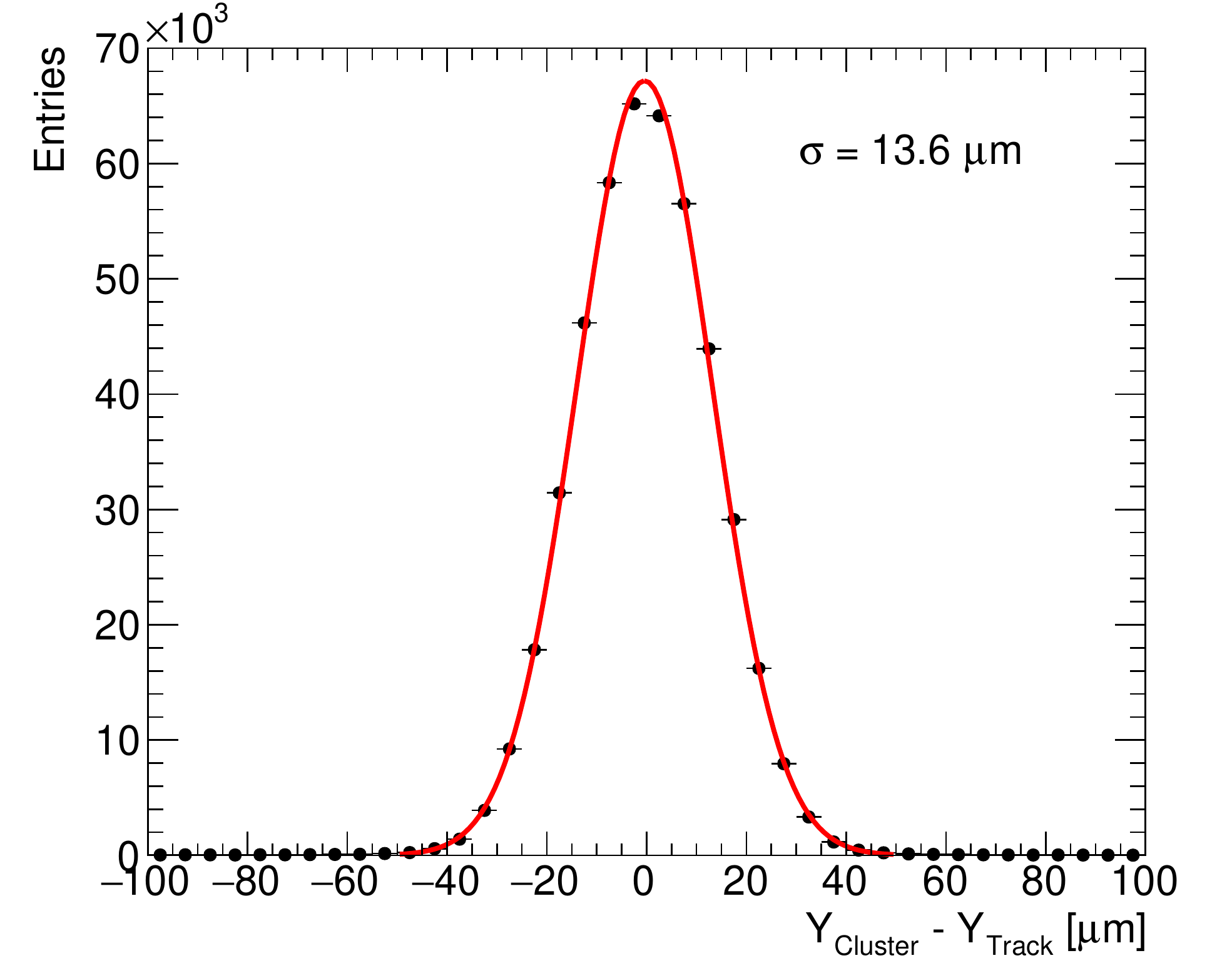}
    \caption{Difference between the expected X- (left) and Y- (right) position of the track in the Mini-MALTA plane and the position of the closest cluster center. The gaussian fit (red lines) yields $\sigma_X=13.5$ $\mu$m and $\sigma_Y=13.6$ $\mu$m.}
    \label{fig:res_xy}
\end{figure}

The residual distribution between the telescope track projection and the center of the nearest cluster, which is not used in the track reconstruction, has a width of $\approx$13.5 $\mu$m in both X and Y directions, as shown in Figure~\ref{fig:res_xy}. The width of the residual distributions is dominated by the expected intrinsic DUT resolution of 10.5 $\mu$m.

\subsection{Hit to track matching and efficiency calculation}
The hit detection efficiency is defined as the fraction of clusters on the DUT matched to telescope tracks over the total number of tracks. Reconstructed tracks with $\chi^2/dof<10$ are used in the efficiency calculation.
A cluster in DUT is matched to a track if the distance between the position of a track interpolated to the plane of the DUT and the position of the center of the cluster is smaller than 100 $\mu$m.

The acceptance area of the DUT is folded into the efficiency calculation as follows: due to the finite track resolution of the beam telescope, all tracks with a hit prediction on a DUT edge pixel are removed from the analysis.
An additional exclusion area of the size of two pixel rows is also applied between the neighboring DUT sensor sectors of different designs. Tracks with hit predictions on noisy pixels (radius of 36.4 $\mu$m around the noisy pixel center) are rejected from the efficiency analysis. Given the limited size of our sensor acceptance in each sector (14$\times$6 pixel after removal of edge pixels) a single noisy pixel can influence the efficiency calculation.


\section{Efficiency measurements before and after irradiation}

\subsection{Efficiency dependency on implant configuration and pre-amplifier gain}

\begin{figure}
    \centering
    \includegraphics[width=.45\textwidth]{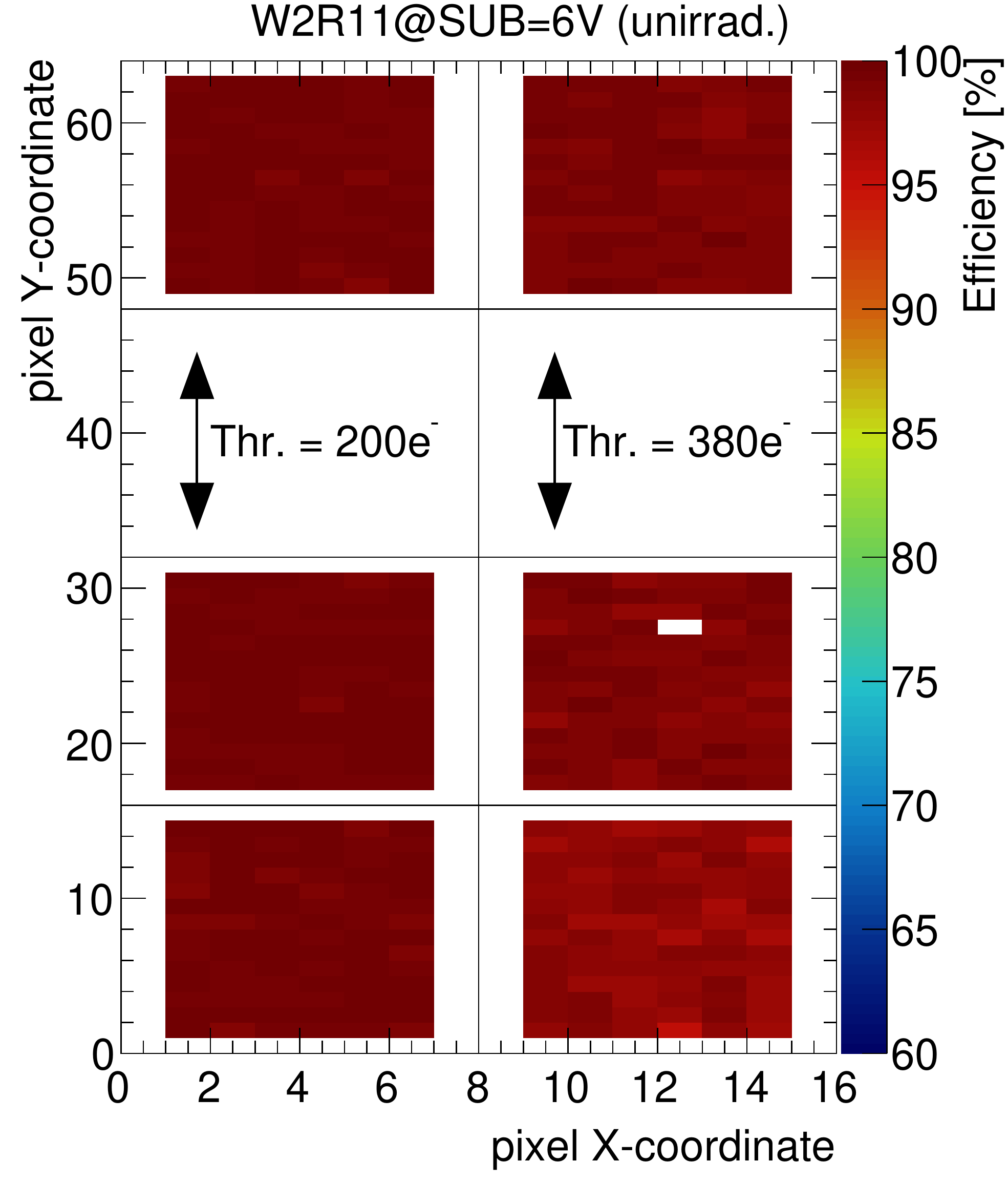}
    \includegraphics[width=.45\textwidth]{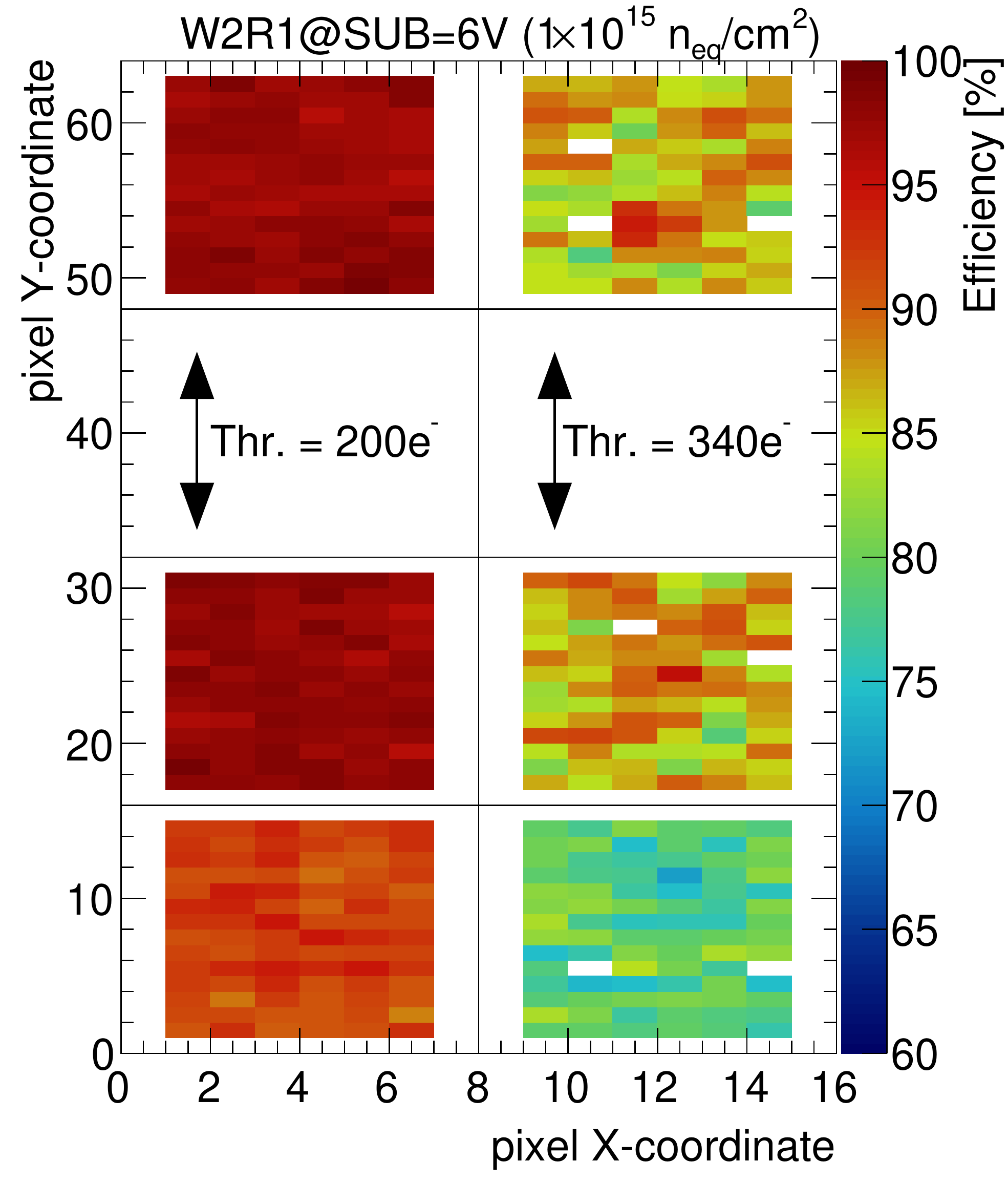}
    \caption{2D efficiency maps for non-irradiated (left) and irradiated Mini-MALTA samples at $1\times10^{15}$ 1 MeV n$_{eq}$/cm$^{2}$ (right).  Different sensor regions are visible: standard MALTA-like (bottom part of each chip), modified with extra deep p-well (middle part) and modified with extra n$^-$ gap (top part). Results are also shown for sensor regions with standard (right side of each chip) and enlarged (left side) transistors. 
    The chips were operated at $-6$ V substrate voltage and $-20^{\circ}$C, and were tuned for low threshold.}
    \label{fig:eff_2d_W2R11_W2R1}
\end{figure}

Figure~\ref{fig:eff_2d_W2R11_W2R1} (left) shows the efficiency as a function of the track position in the DUT plane for an unirradiated Mini-MALTA sensor (``W2R11'') tuned to an example threshold of 200e$^{-}$ on sectors with enlarged transistors and 380e$^{-}$ on sectors with standard, i.e. minimum size, transistors. The measured sector efficiency for an unirradiated chip and standard transistors were 97.9$\pm$0.1\% for the continuous n-layer, 98.9$\pm$0.1\% for the extra deep p-well and 99.1$\pm$0.1\% for the n$^-$ gap.

For sensor regions with enlarged transistors the average efficiency is 99.6$\pm$0.1\% at a threshold of 200e$^{-}$ and does not depend on extra sensor modification. 
The small inefficiency is partly due to relatively tight cluster-to-track matching conditions which accounts for 0.2\% efficiency loss, and partly due to small regional inefficiency at double-column boundaries ($\approx$0.2\% overall efficiency reduction) which is still under investigation. 

The 2D efficiency map for a $1\times10^{15}$ 1 MeV n$_{eq}$/cm$^{2}$ neutron irradiated Mini-MALTA sensor (``W2R1'') is shown in Figure~\ref{fig:eff_2d_W2R11_W2R1} (right). 

Table~\ref{tab:effi} lists the measured efficiency at given threshold values for all Mini-MALTA sensors used in the beam tests in dependency of the implant configuration and FE amplifier design. 

\begin{table}[h!]
\centering
\begin{tabular}{ |C{1cm}|C{0.7cm}|C{2.2cm}|C{0.7cm}|C{2.5cm}|C{3cm}|C{3cm}| }
\hline
Chip ID & EPI [$\mu$m] & Fluence [1 MeV n$_{eq}$/cm$^{2}$] & SUB [V] & Process modification & Efficiency (enlarged trans. region) [\%] / & Efficiency (standard trans. region) [\%] / \\
 & & & & & threshold [e$^{-}$] & threshold [e$^{-}$]  \\
\hline \hline
\multirow{3}{*}{W2R11} &\multirow{3}{*}{30} & \multirow{3}{*}{unirrad.} & \multirow{3}{*}{$-6$}
 &     n$^-$ gap          & $99.6\pm 0.1$ / 200e$^{-}$& $99.1\pm 0.1$ / 380e$^{-}$\\
& & & & extra deep p-well    & $99.6\pm 0.1$ / 200e$^{-}$& $98.9\pm 0.1$ / 380e$^{-}$\\
& & & & continuous n$^{-}$    & $99.6\pm 0.1$ / 200e$^{-}$& $97.9\pm 0.1$ / 380e$^{-}$\\ \hline
          
\multirow{3}{*}{W2R1} &\multirow{3}{*}{30} & \multirow{3}{*}{$1\times10^{15}$} & \multirow{3}{*}{$-6$}
 &      n$^-$ gap          & $97.6\pm 0.1$ / 105e$^{-}$& $86.5\pm 0.1$ / 210e$^{-}$\\
& & & & extra deep p-well    & $97.9\pm 0.1$ / 105e$^{-}$& $87.0\pm 0.1$ / 210e$^{-}$\\
& & & & continuous n$^{-}$  & $91.9\pm 0.1$ / 105e$^{-}$& $78.8\pm 0.2$ / 210e$^{-}$\\ \hline

\multirow{3}{*}{W4R2} &\multirow{3}{*}{25} & \multirow{3}{*}{$1\times10^{15}$} & \multirow{3}{*}{$-6$}
 &      n$^-$ gap          & $98.8\pm 0.1$ / 120e$^{-}$& $90.7\pm 0.1$ / 275e$^{-}$\\
& & & & extra deep p-well    & $99.2\pm 0.1$ / 120e$^{-}$& $92.5\pm 0.1$ / 275e$^{-}$\\
& & & & continuous n$^{-}$   & $95.8\pm 0.1$ / 120e$^{-}$& $79.4\pm 0.2$ / 275e$^{-}$\\ \hline

\multirow{3}{*}{W5R3} &\multirow{3}{*}{25} & \multirow{3}{*}{$2\times10^{15}$} & \multirow{3}{*}{$-10$}
 &      n$^-$ gap          & $92.1\pm 0.2$ / 120e$^{-}$& $73.1\pm 0.3$ / 230e$^{-}$\\
& & & & extra deep p-well    & $93.7\pm 0.2$ / 120e$^{-}$& $76.4\pm 0.3$ / 230e$^{-}$\\
& & & & continuous n$^{-}$   & $86.5\pm 0.2$ / 120e$^{-}$& $70.9\pm 0.3$ / 230e$^{-}$\\ 
\hline
\hline
\end{tabular}
\caption{Summary of the efficiency measurements for various Mini-MALTA chips. The values are shown separately for different sensor regions. 
All chips were operated at low threshold settings. The uncertainties listed are statistical.}
\label{tab:effi}
\end{table}

\begin{figure}
    \centering
    \includegraphics[width=.45\textwidth]{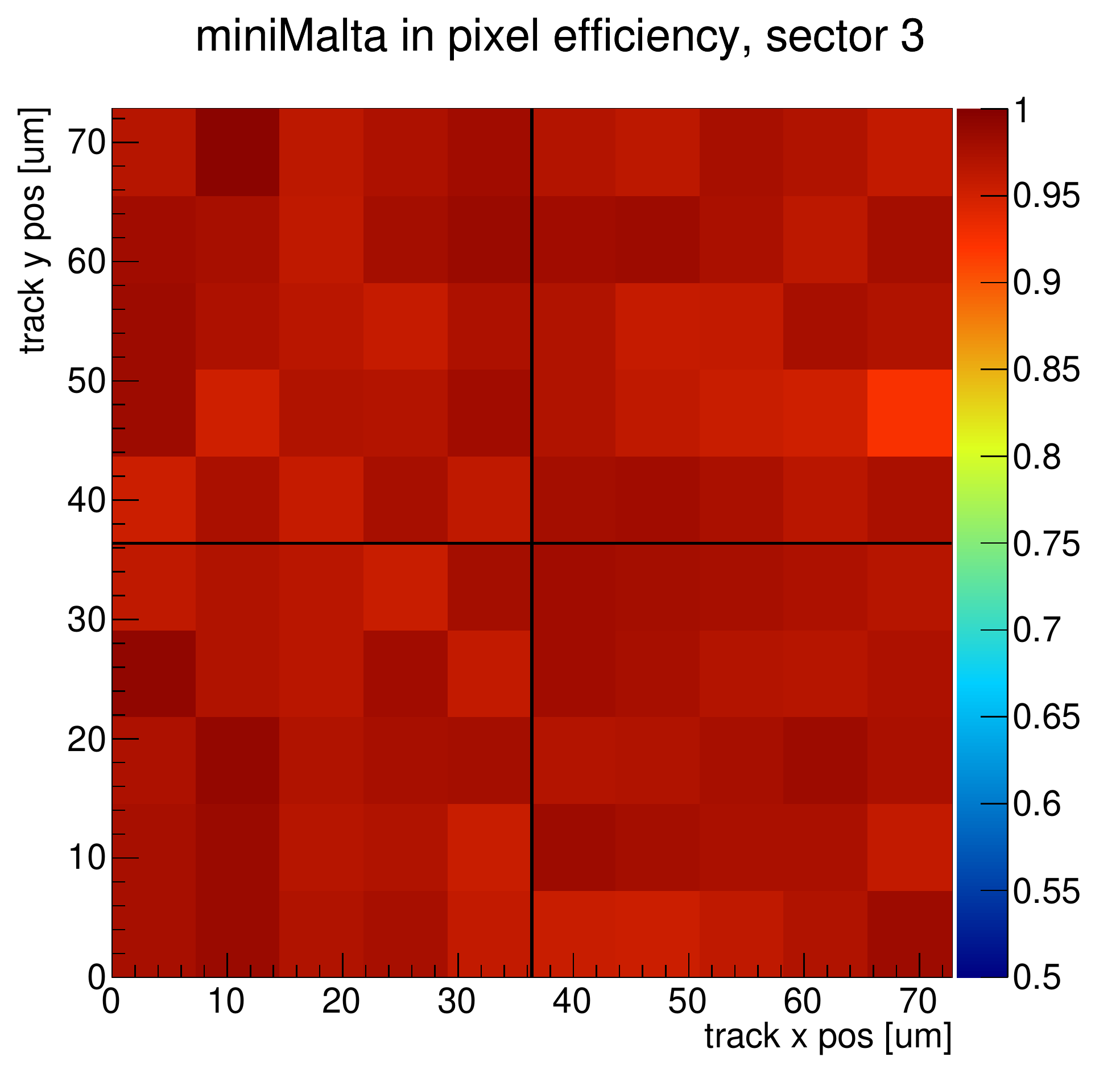}
    \includegraphics[width=.45\textwidth]{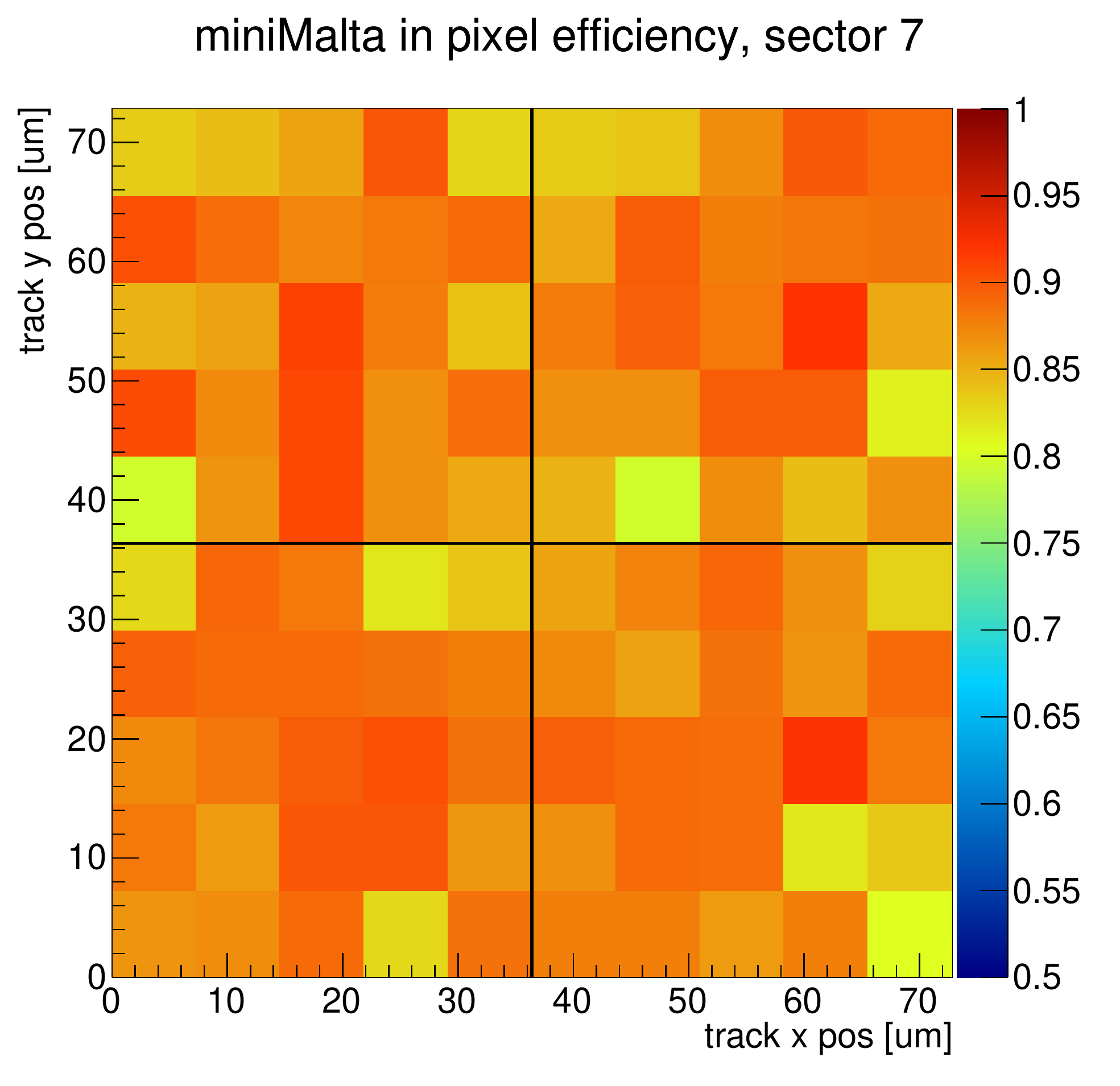}
    \includegraphics[width=.45\textwidth]{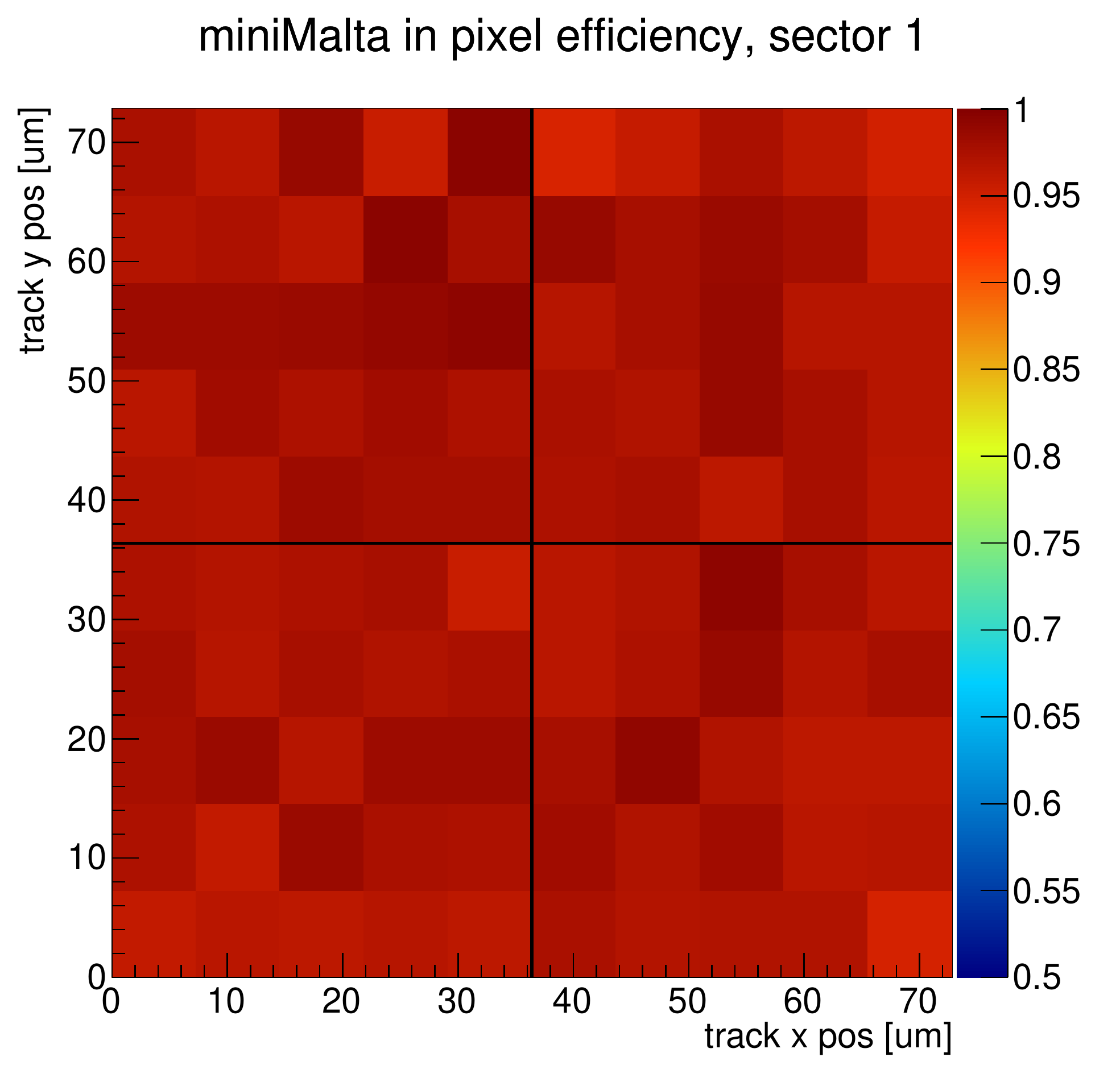}
    \includegraphics[width=.45\textwidth]{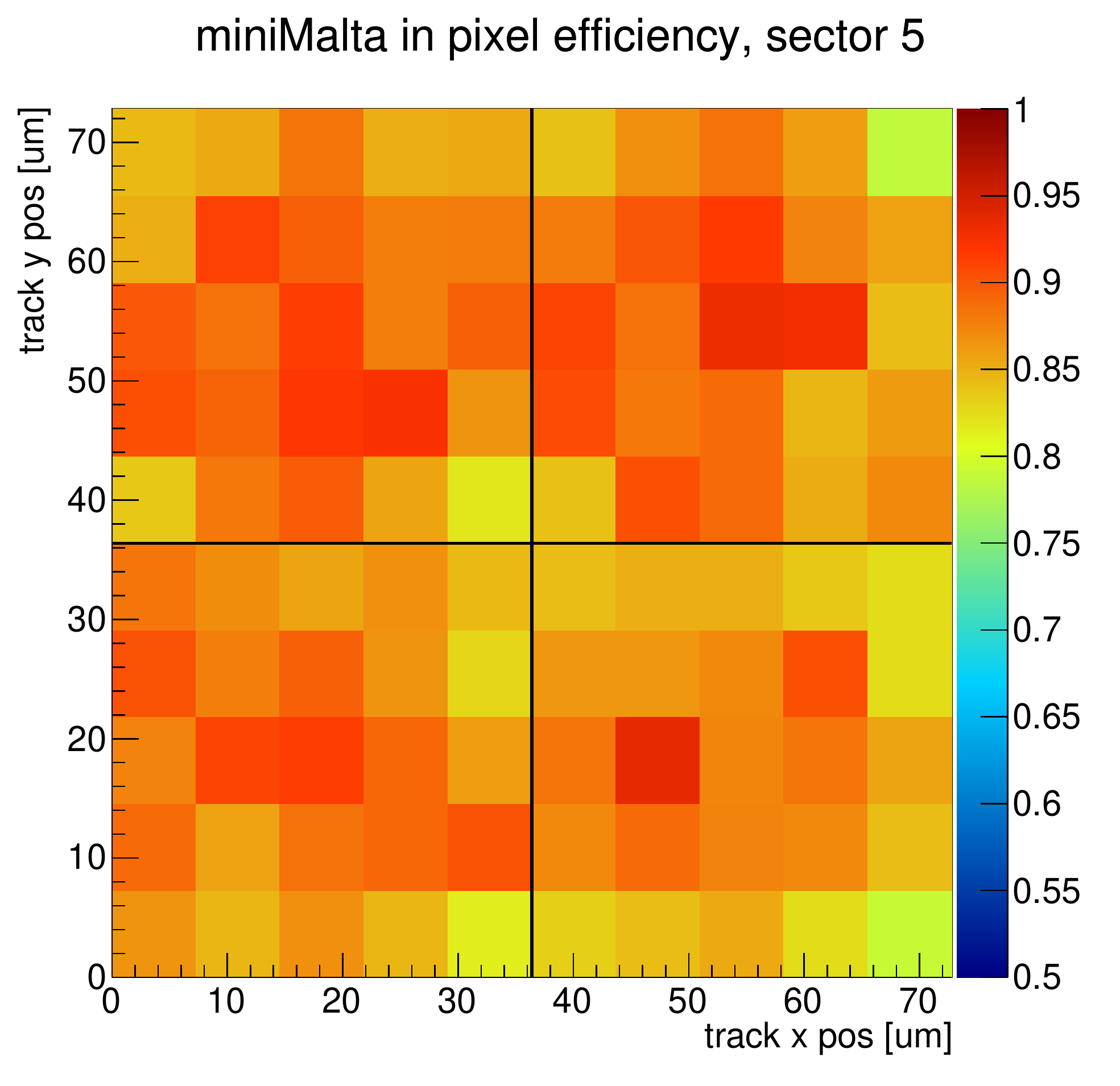}
    \includegraphics[width=.45\textwidth]{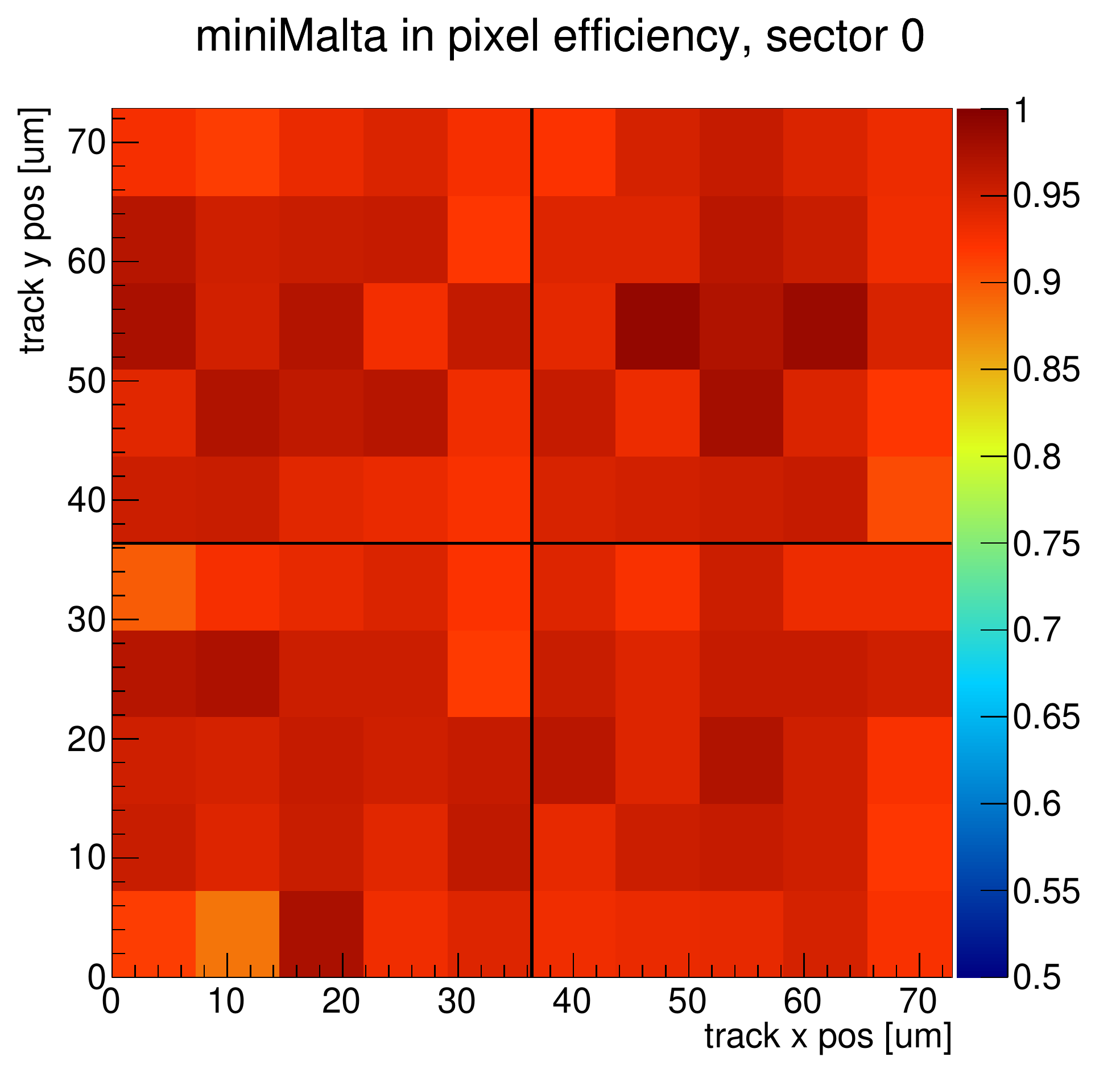}
    \includegraphics[width=.45\textwidth]{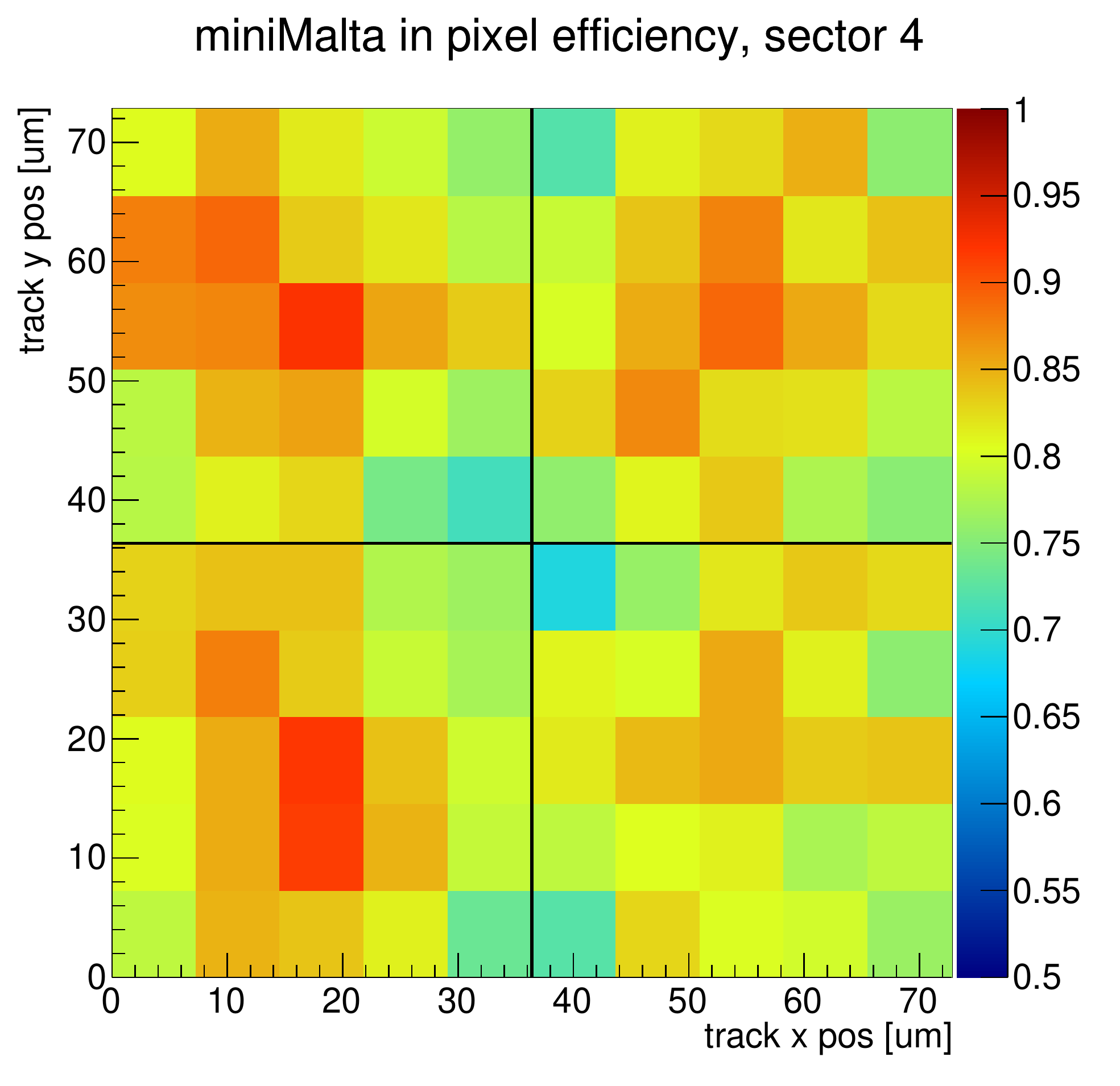}
    \caption{In-pixel 2D efficiency maps for irradiated Mini-MALTA sample at $1\times10^{15}$ 1 MeV n$_{eq}$/cm$^{2}$ for $2\times2$ pixel groups. Different sensor regions are shown: standard MALTA-like (bottom part of each chip), modified with extra deep p-well (middle part) and modified with extra n$^-$ gap (top part). Results are also shown for sensor regions with standard (right side of each chip) and enlarged (left side) transistors. 
    The binning corresponds to $5\times5$ entries per single pixel.
    The chip was operated at $-6$ V substrate voltage and $-20^{\circ}$C.}
    \label{fig:inpix_eff_W2R1_1e15}
\end{figure}

After $1\times10^{15}$ 1 MeV n$_{eq}$/cm$^{2}$ neutron irradiation, the average efficiency significantly decreases for regions with standard transistors due to the lower gain (higher effective threshold) in these sectors: it reaches 78.8\% in the region with no extra sensor modification, 87.0\% with extra deep p-well and 86.5\% with n$^-$ gap. While the modifications to the implant proof effective, the overall efficiency is still significantly affected by the high threshold.
As expected, the inefficiency is mainly present in regions around the pixel edges and corners, as shown in Figure~\ref{fig:inpix_eff_W2R1_1e15} where the in-pixel efficiency plots are shown. 
The 2$\times$2 pixel plots overlap with the double column structure of the chip matrix. Along the pixel edge and in particular in the corners the charge is shared between two or more pixels. The resulting small per-pixel signals are suppressed by a high effective threshold leading to inefficiency.
For regions with enlarged transistors the average efficiency is 91.9\% in the region with no extra sensor modification, 97.9\% with extra deep p-well and 97.6\% with extra n$^-$ gap.
It is therefore clear that the usage of enlarged transistors significantly increases efficiency for sensors after irradiation due to the higher gain, lower gain spread and reduced RTS noise. In the implant configuration with a continuous n$^{-}$ layer and no extra deep p-well (sector 0) the overall efficiency is still reduced in the corner. If the charge collection in the pixel corners is improved through either a gap in the n$^{-}$ layer or an extra deep p-well along the pixel borders (see Figure~\ref{fig:process_modification} bottom row), the sensor becomes nearly full efficient at $1\times10^{15}$ 1 MeV n$_{eq}$/cm$^{2}$. 

Figure~\ref{fig:eff_2d_W2R11_W2R1} also shows the number of noisy pixels through white bins in the different sectors, where the track prediction is rejected and no efficiency is calculated. We observe on the irradiated sensor (``W2R1'') noisy pixels only in the area of standard FE. In the area with enlarged transistors, no noisy pixels are visible. This provides another indication that enlarging some crucial NMOS transistors helps to reduce RTS noise. 

\begin{figure}
    \centering
    \includegraphics[width=.49\textwidth]{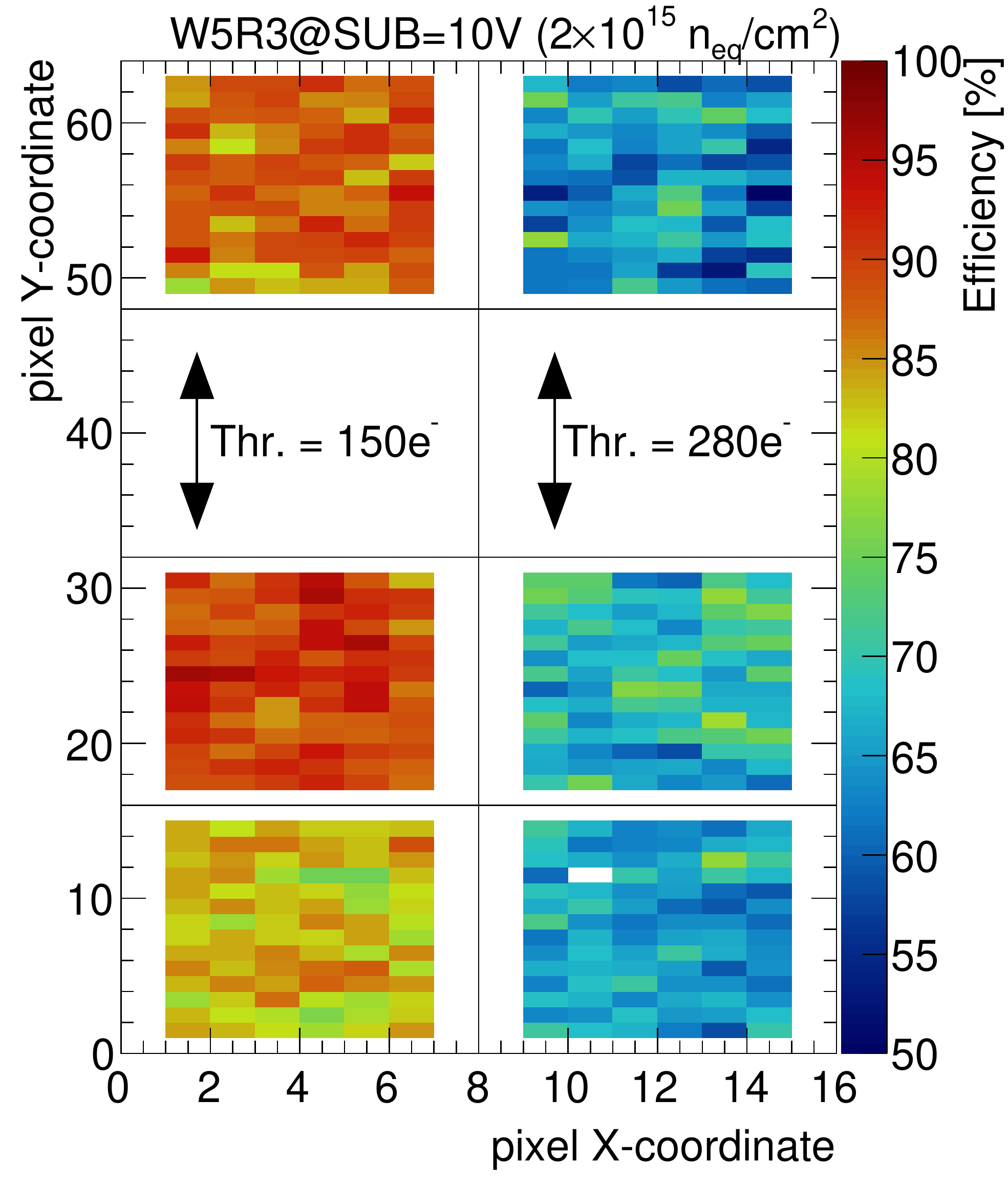}
    \includegraphics[width=.49\textwidth]{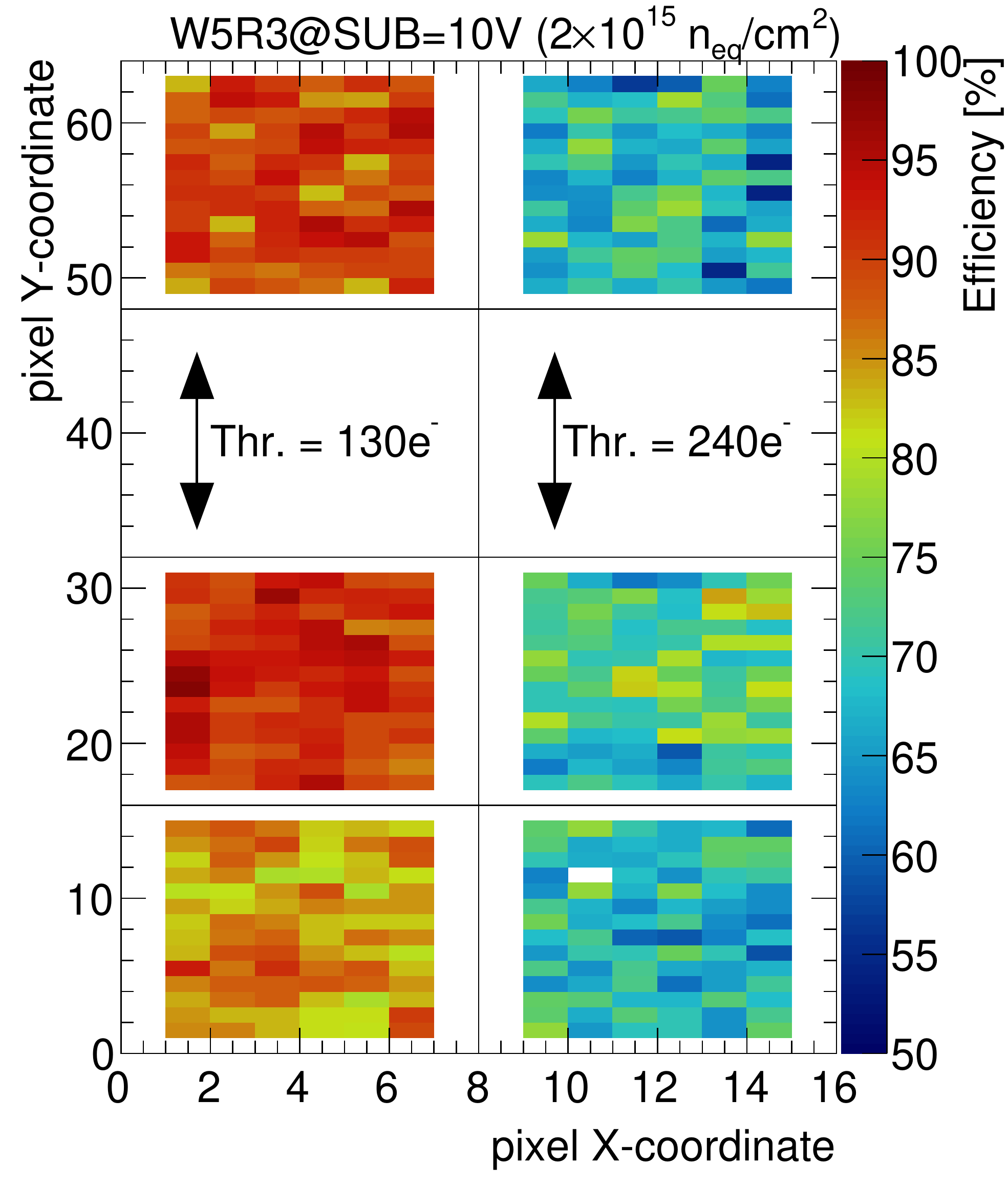}
    \includegraphics[width=.49\textwidth]{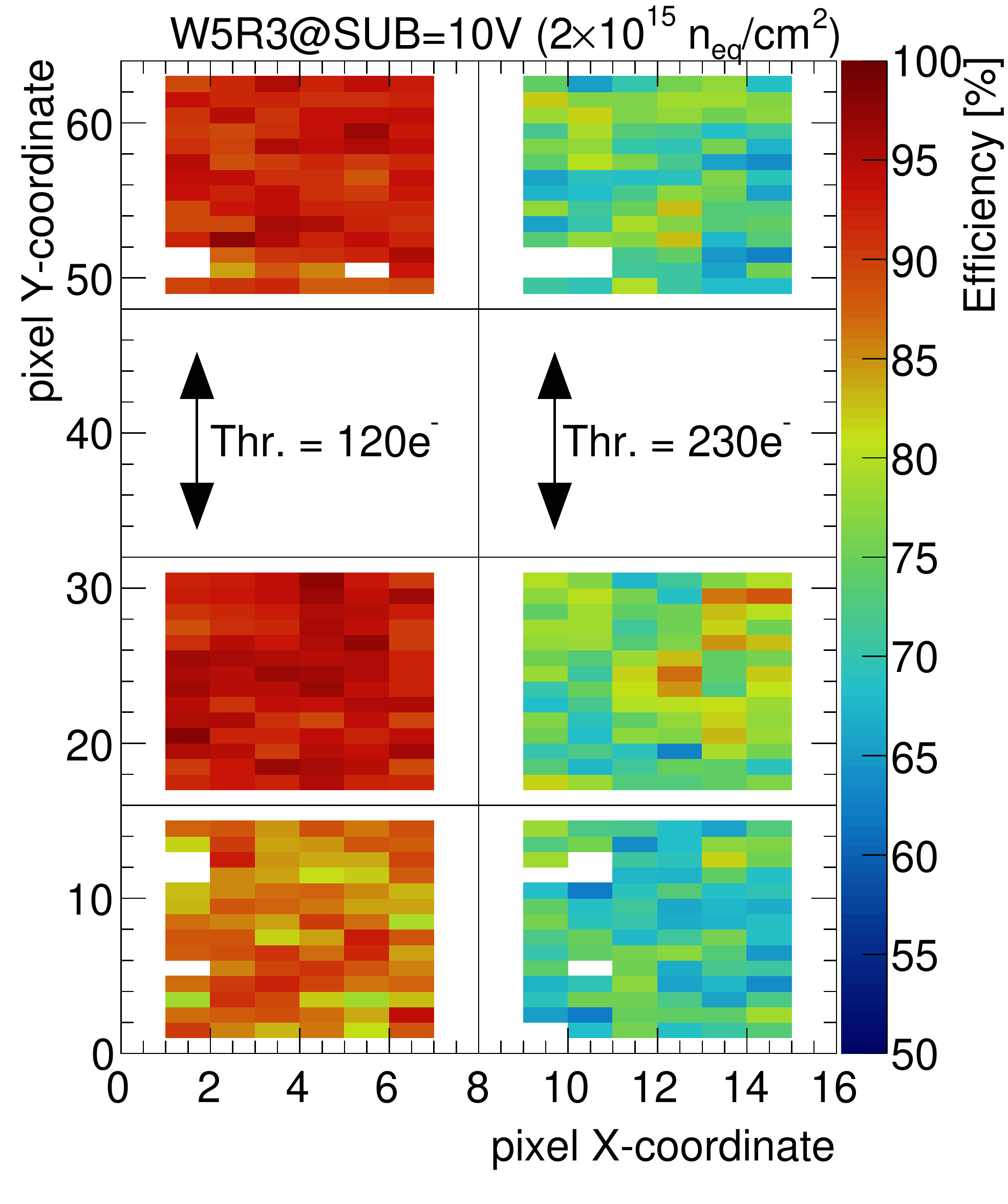}
    \caption{2D efficiency maps for irradiated Mini-MALTA sample at $2\times10^{15}$ 1 MeV n$_{eq}$/cm$^{2}$ at various threshold settings. Different sensor regions are visible: standard MALTA-like (bottom part of each chip), modified with extra deep p-well (middle part) and modified with extra n$^-$ gap (top part). Results are also shown for sensor regions with standard (right side of each chip) and enlarged (left side) transistors. 
    The binning corresponds to one entry per single pixel.
    The chip was operated at $-10$ V substrate voltage and $-20^{\circ}$C.}
    \label{fig:eff_2d_2e15}
\end{figure}

After $2\times10^{15}$ 1 MeV n$_{eq}$/cm$^{2}$ neutron irradiation, shown for sensor ``W5R3'' at different thresholds in Figure~\ref{fig:eff_2d_2e15}, we observe an efficiency reduction to $\approx$93\% also in the sector with extra deep p-well and $\approx$92\% in the sector with n$^-$ gap. All other sectors are affected by this efficiency reduction as well. The higher charge thresholds further degrade the efficiencies in the sectors with the lower gain pre-amplifier, especially after irradiation.

\subsection{Efficiency dependence on substrate voltage}
The efficiency is also studied as function of the substrate voltage. From TCAD simulations we expect best charge collection at a substrate voltage of around $-10$ V~\cite{Munker:2019vdo}. Higher substrate voltage leads to a strong vertical electric field but reduces the relative lateral field. The result of a higher substrate voltage is a much more vertical field that pushes charge generated near the pixel corners/edges into the low field region under the p-well at those locations. The charge takes time to escape from these regions and this decreases efficiency especially after irradiation as then charge is more easily captured by radiation induced traps. 

\begin{figure}
    \centering
    \includegraphics[width=.49\textwidth]{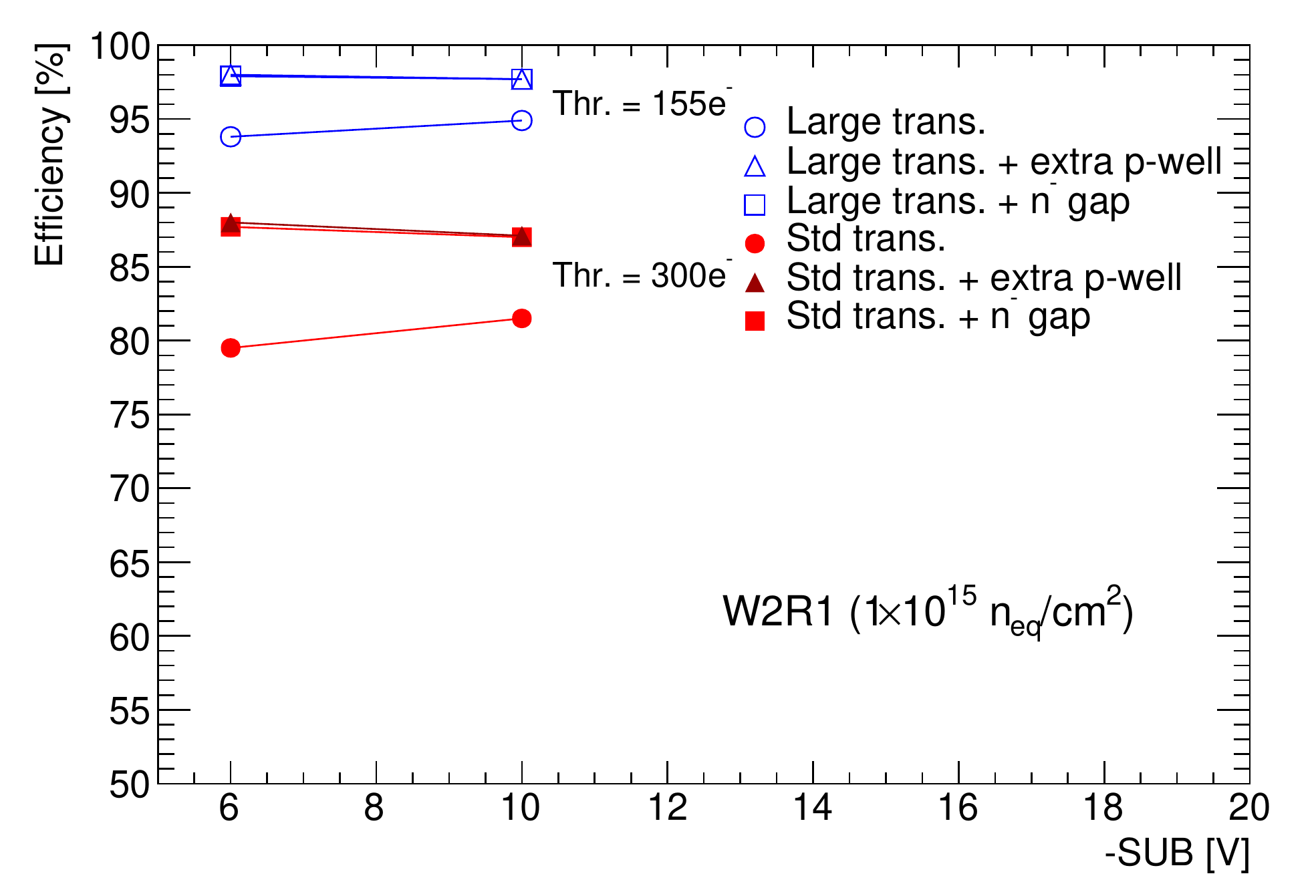}
    \includegraphics[width=.49\textwidth]{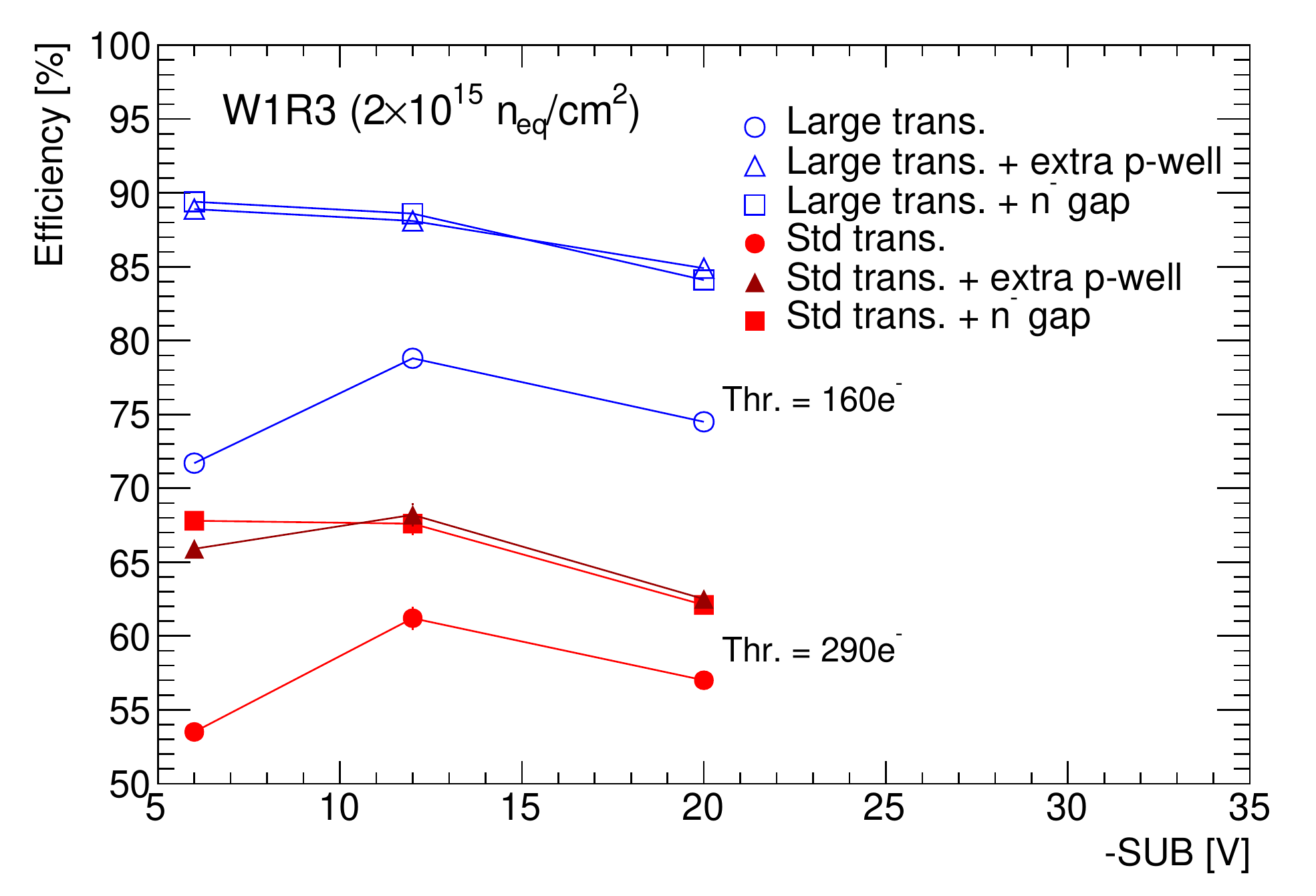}
    \caption{Efficiency versus SUB voltage for neutron irradiated Mini-MALTA samples at $1\times10^{15}$ 1 MeV n$_{eq}$/cm$^{2}$ (left) and $2\times10^{15}$ 1 MeV n$_{eq}$/cm$^{2}$ (right).  Different sensor regions are presented: standard MALTA-like (circles), modified with extra deep $p$-well (triangles) and modified with extra n$^-$ gap (rectangles). Results are also shown for sensor regions with standard (open markers) and enlarged (full markers) transistors.}
    \label{fig:eff_vs_sub_1e15_2e15}
\end{figure}

In Figure~\ref{fig:eff_vs_sub_1e15_2e15} we show the results for neutron irradiated Mini-MALTA samples at $1\times10^{15}$ and  $2\times10^{15}$ 1 MeV n$_{eq}$/cm$^{2}$.
For the chip irradiated at $1\times10^{15}$ 1 MeV n$_{eq}$/cm$^{2}$ the efficiency is relatively stable when changing the substrate voltage between $-6$ V and $-10$ V.
In contrast, the chip irradiated at $2\times10^{15}$ 1 MeV n$_{eq}$/cm$^{2}$ has best efficiency around -10~V to -12~V. The efficiency decreases at higher ($-6$ V) and lower ($-20$ V) SUB voltages in most of the sectors which confirms the qualitative observations in TCAD simulations.
Therefore, the non-irradiated and irradiated Mini-MALTA samples at $1\times10^{15}$ 1 MeV n$_{eq}$/cm$^{2}$ were operated at a substrate voltage of $-6$~V, whereas the samples irradiated at $2\times10^{15}$ 1 MeV n$_{eq}$/cm$^{2}$ had a substrate voltage of $-10$ V applied.

\subsection{Efficiency for different threshold settings}
\begin{figure}
    \centering
    \includegraphics[width=.7\textwidth]{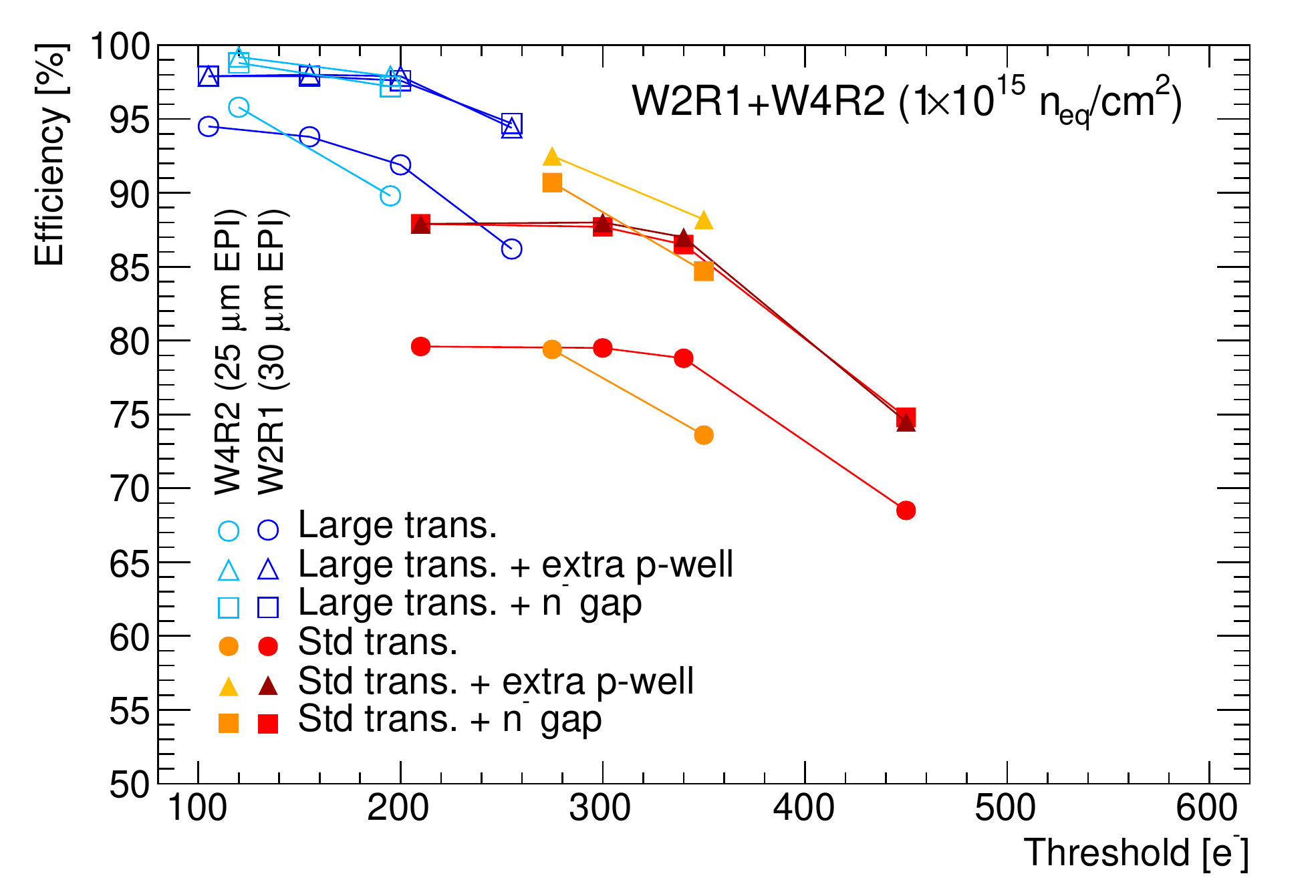}
    \caption{Efficiency versus threshold mean for neutron irradiated Mini-MALTA samples at $1\times10^{15}$ 1 MeV n$_{eq}$/cm$^{2}$ ($-6$ V substrate voltage, $-20^{\circ}$C). Different sensor regions are presented: standard MALTA-like (circles), modified with extra deep p-well (triangles) and modified with extra n$^-$ gap (rectangles). Results are also shown for sensor regions with standard (open markers) and enlarged (full markers) transistors, as well as for sensors with different epitaxial layer thicknesses: 25 $\mu$m (orange or light blue symbols) and 30 $\mu$m (red or dark blue symbols).}
    \label{fig:eff_vs_th_1e15}
\end{figure}

The efficiency is significantly affected by the preamplifier design, and here in particular the size of the NMOS transistor ``M3'' influencing gain, gain uniformity and RTS noise, ultimately allowing lower threshold operation to achieve higher efficiency.
Furthermore, the additional deep p-type implant or n$^-$ gap at the pixel edges improves the charge collection and efficiency after irradiation. The efficiency as a function of threshold for $1\times10^{15}$ 1 MeV n$_{eq}$/cm$^{2}$ irradiated Mini-MALTA samples is shown in Figure~\ref{fig:eff_vs_th_1e15}.
With lowering the threshold, the efficiency increases and can reach approximately 95\% for continuous n$^{-}$ sectors with enlarged transistors and 98--99\% for sectors with n$^-$ gap or extra deep p-well. The gain is due to better charge collection in the pixel corners caused by the modifications of implants on the pixel edge.

When using minimal size transistors we reach only 88--92\% efficiency for n$^-$ gap or extra deep p-well pixel designs even at thresholds around 200~e$^{-}$ whereas the preamplifier design with enlarged transistors reaches 98\%. This highlights the importance of transistor choice in our pre-amplifier circuit, most notably the ``M3''transistor which needs to be chosen for lowest output conductance. The sector with continuous n$^{-}$ and standard transistor behaves worst, due to the inefficient charge collection in the pixel corners and high threshold.

\begin{figure}
    \centering
    \includegraphics[width=.7\textwidth]{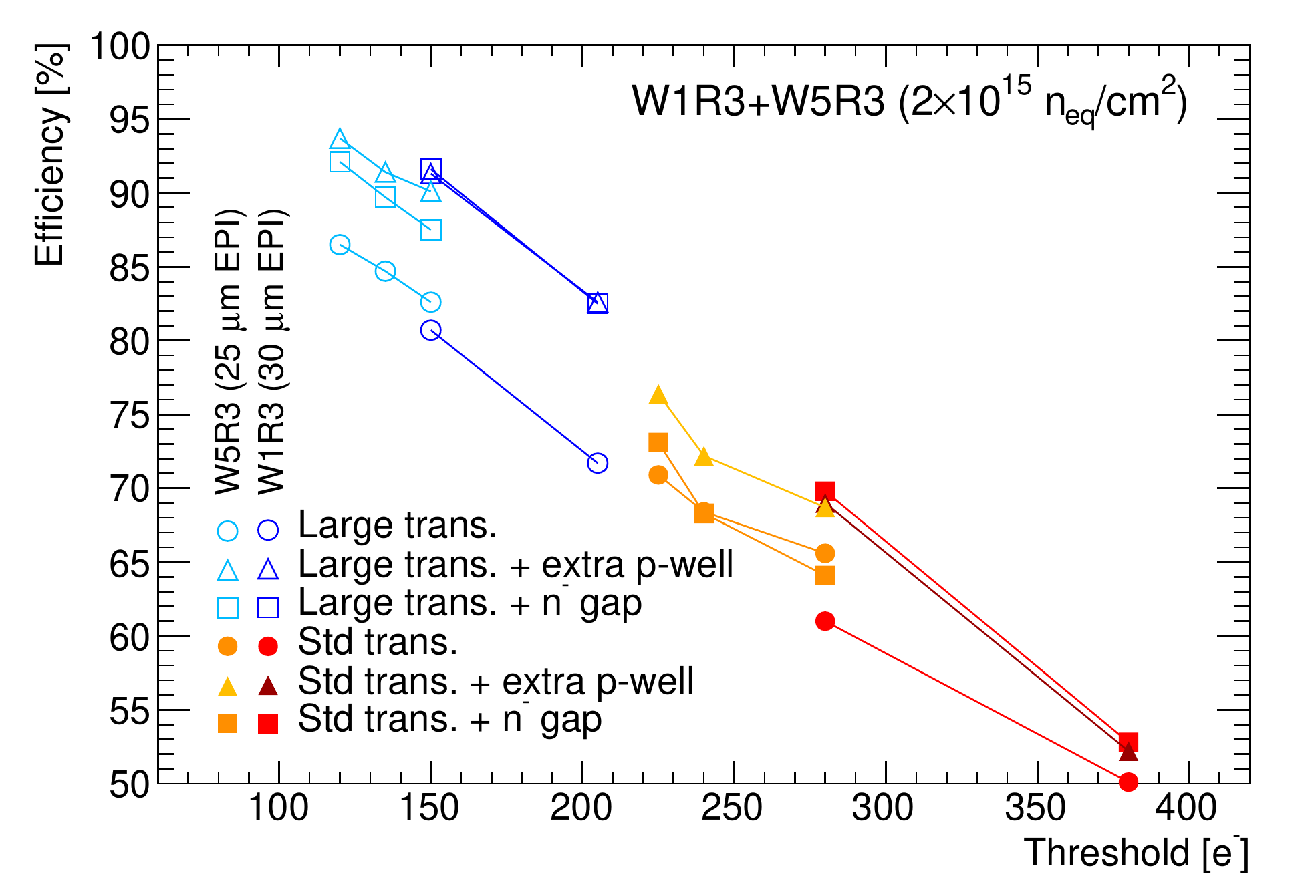}
    \caption{Efficiency versus threshold mean for neutron irradiated Mini-MALTA samples at $2\times10^{15}$ 1 MeV n$_{eq}$/cm$^{2}$ ($-10$ V substrate voltage, $-20^{\circ}$C). Different sensor regions are presented: standard MALTA-like (circles), modified with extra deep p-well (triangles) and modified with extra n$^-$ gap (rectangles). Results are also shown for sensor regions with standard (open markers) and enlarged (full markers) transistors, as well as for sensors with different epitaxial layer thicknesses: 25 $\mu$m (orange or light blue symbols) and 30 $\mu$m (red or dark blue symbols).}
    \label{fig:eff_vs_th_2e15}
\end{figure}

Similar trends are visible for neutron irradiated Mini-MALTA samples at $2\times10^{15}$ 1 MeV n$_{eq}$/cm$^{2}$ (Figure~\ref{fig:eff_vs_th_2e15}).
In this case the best efficiency is obtained for modified regions with extra deep $p$-well and amounts up to 75\% for regions with standard-size transistors and up to 94\% for regions with enlarged transistors.

In these measurements we also observe that sensors with 25~$\mu$m epitaxial layer produce similar or sightly better efficiency than sensors produced on 30 $\mu$m epitaxial wafers at the same substrate voltage, despite the fact that we expect about 20\% more ionization charge in the 30 $\mu$m epitaxial wafers. This effect may be due to the higher field in the thinner epitaxial layer. Additionally the choice of processing for the creation of the low-doped n$^{-}$ layer may contribute to the observed behaviour: For sensors produced on 25  $\mu$m epitaxial wafers we have chosen a slightly deeper n$^{-}$ implantation than in the 30 $\mu$m epitaxial wafers. A deeper n$^{-}$ layer moves the junction deeper in the high resistivity epitaxial layer which provides a better field configuration for charge collection in the pixel corners. 

\section{Conclusions}


This paper presents measurement results on the Mini-MALTA Monolithic Active Pixel Sensor prototype developed in the TowerJazz 180nm CMOS imaging process. The prototype implements several improvements to address the inefficiencies and limitations of the previous prototypes. In particular charge collection is improved in the pixel corners by modifying implants along the pixel boundary (n$^-$ gap or extra deep p-well pixel designs), and the charge sensitive FE is improved by increasing the size of an NMOS transistor yielding lower RTS noise, higher gain and lower threshold spread allowing operation at lower thresholds. 

The measurement results demonstrate that with these improvements the sensors achieve a 98-99 \% efficiency at a threshold of 100e$^{-}$ to 150e$^{-}$ after a dose of $1\times10^{15}$ 1 MeV n$_{eq}$/cm$^{2}$. To achieve full efficiency at $2\times10^{15}$ 1 MeV n$_{eq}$/cm$^{2}$ the FE should be further improved to achieve thresholds below 100e$^{-}$.



\section{Acknowledgements}
The authors are grateful to the University of Bonn for the support received during measurements performed at the E3 beam-line at the electron accelerator ELSA operated by the university of Bonn in Nordhrein-Westfalen, Germany. We are also grateful to the Institute Jo\v{z}ef Stefan, Ljubljana, Slovenia during the irradiation for the support during neutron irradiations. The irradiation campaign has been supported by the H2020 project AIDA-2020, GA no. 654168. The beam test measurements have also received support by the Turkish Atomic Energy Authority (TAEK) under the project grant no. 2018TAEK(CERN)A5.H6.F2-20. This research project has been supported by the Marie Sklodowska-Curie Innovative Training Network of the European Commission Horizon 2020 Programme under contract number 675587 ``STREAM''.

\bibliographystyle{JHEP}
\bibliography{main}

\end{document}